\documentclass[aps,pra.reprint,twocolumn]{revtex4} 
\usepackage{natbib}
\usepackage{epsfig}
\usepackage{graphicx}
\usepackage{amsmath}
\usepackage{amsfonts}
\usepackage{amssymb}
\usepackage{cancel}
\usepackage{graphicx}
\usepackage{ulem}
\usepackage{wrapfig}
\usepackage[cp1251]{inputenc}
\usepackage{epsfig}
\usepackage{graphicx}
\usepackage{amsmath}
\usepackage{amsfonts}
\usepackage{amssymb}
\usepackage{graphicx}
\usepackage{wrapfig}
\usepackage{textcomp}
\usepackage{filecontents}
\usepackage{pgf,pgfarrows,pgfnodes,pgfautomata,pgfheaps,pgfshade}
\usepackage[colorlinks=true, linkcolor=blue]{hyperref}
\usepackage{lipsum}

\begin{document}

\title{
Universal fluctuations and squeezing  in generalized Dicke model near the superradiant phase transition}

\author{D. S. Shapiro$^{1,2,3}$} \email{shapiro.dima@gmail.com} 	
	\author{W. V. Pogosov$^{1,4}$}
	 \author{Yu. E. Lozovik$^{5,1,6}$}  
	\affiliation{$^1$Dukhov Research Institute of Automatics (VNIIA),   127055 Moscow, Russia}
	\affiliation{$^2$Department of Physics, National Research University Higher School of Economics, 101000 Moscow, Russia}
 \affiliation{ 	 $^3$Laboratory of Superconducting Metamaterials, National University of Science and Technology MISiS,   119049 Moscow, Russia}  
	\affiliation{$^4$Institute for Theoretical and Applied Electrodynamics, Russian Academy of
		Sciences, 125412 Moscow, Russia}
\affiliation{$^5$Institute of Spectroscopy, Russian Academy of Sciences, 142190 Moscow region,
	Troitsk, Russia}
\affiliation{$^6$Moscow Institute of Physics and Technology, Dolgoprudny, Moscow Region 141700, Russia
	}

\begin{abstract}
In a view of recent proposals for the realization of anisotropic light-matter interaction in  such   platforms as (i) non-stationary or inductively and capacitively coupled superconducting qubits,  (ii) atoms in crossed fields and (iii)  semiconductor heterostructures with spin-orbital interaction, the concept of  generalized Dicke model, where  coupling strengths of rotating wave and counter-rotating wave terms are unequal, has attracted   great interest. For this model, we   study   photon fluctuations   in the critical region of normal-to-superradiant phase transition   when both the temperatures and  numbers of two-level systems are   finite.   In this case, the  superradiant quantum phase transition is changed to  a  fluctuational region in the phase diagram that reveals two types of  critical behaviors. These are  regimes of  Dicke model (with discrete $\mathbb{Z}_2$ symmetry), and that of  (anti-) and  Tavis-Cummings  $U(1)$ models.
   We show that   squeezing parameters of photon condensate  in these regimes show  distinct temperature scalings. Besides,  relative fluctuations of  photon number take universal values. 
   We also find a  temperature scales below which one approaches  zero-temperature  quantum phase transition where  quantum fluctuations dominate. Our effective theory is provided by a non-Goldstone functional for  condensate mode  and  by Majorana representation of Pauli operators.  We also discuss Bethe ansatz solution for     integrable $U(1)$ limits.

\end{abstract}

\maketitle

\section{Introduction}

  An important concept of contemporary quantum optics and   
cavity quantum electrodynamics is  a  single mode version of Dicke model~\cite{PhysRev.93.99}, where an ensemble of two-level systems   interacts with    quantized    electromagnetic field in a cavity, microwave resonator, etc. 
This model demonstrates   superradiant phase transition,  a collective  
phenomenon   characterized by a condensation of macroscopic number of photons.
 Experimental signatures of 2$^{\rm nd}$ order  quantum phase transition, equivalent to the superradiant one, were observed in a driven Bose-Einstein condensate of Rb atoms in an optical cavity~\cite{Baumann}. Also, the engineering of the Dicke model simulator with cold Be atoms in optical trap and signatures of  superradiant phase transition  were reported in~\cite{PhysRevLett.121.040503}.
 The physics of the Dicke  model  is believed to be tested in quantum metamaterials such as   superconducting qubits arrays \cite{macha2014implementation, PhysRevLett.117.210503, Shulga2017, Zhang2017}   integrated with a GHz transmission line via tunable couplers \cite{Srinivasan2011,Hoffman2011,Chen2014,Zeytinouglu2015}. The recent advances in implementations of strong coupling regimes in superconducting circuits~\cite{PhysRevLett.105.237001, Bosman2017,Andersen_2017,braumuller2017analog}  are promising for realizations of phase transitions as well. Extremely fast emission, indicating for   a superradiant pulse, was observed in  lumped  resonator   coupled to an inhomogeneously broadened macroscopic ensemble of nitrogen-vacancy centers~\cite{Putz2014,Angerer2018}.

 	Thank to advances in fabrication  technologies of light-matter hybrid systems during last years,  an interest  to generalizations  of the Dicke model has emerged.
 	 	The behavior in a presence of incoherent pumping  or cavity loss   reveals a richness of phase diagrams, see Ref.~\cite{kirton2018introduction} for a review.
 	In the present work we are focused on another example of generalization, the anisotropic  qubit-cavity  interaction, i.e., when strengths of rotating- and counter-rotating wave terms  are different.
	The possible      physical realizations   are frequency-modulated~\cite{Wang_freq_mod_qrm} or inductively and capacitively coupled~\cite{PhysRevLett.112.173601}  superconducting qubits,  semiconductor heterostructures with spin-orbital interaction~\cite{Wang_QRM_spin-orbit} and   atoms in crossed electric and magnetic fields,  see Ref.\cite{PhysRevX.4.021046} for a review and  also references therein.

 The  Hamiltonian of  the generalized  Dicke model (GDM) reads as
 \begin{multline}
 	\hat H=  \omega\hat a^\dagger\hat a +\frac{\epsilon }{2}\hat S_z+	\\ +\frac{g}{\sqrt N}(  \hat a \hat S^+ + \hat a^\dagger \hat S^-  )     +\frac{J}{\sqrt N}(  \hat a \hat S^- + \hat a^\dagger \hat S^+  )  \  . \label{h}
 \end{multline}
 The first term describes single-mode photon field of the excitation  frequency $\omega$; here $\hat a^\dagger$ and $\hat a$ are   respective creation and annihilation operators. The second term is the Hamiltonian of  the ensemble consisting of $N$ two-level systems.   
They have  equal energy splittings $\epsilon$  between   their ground and excited states. The collective  angular momentum operator $\hat S^{z}=\sum\limits_{j=1}^N   \hat\sigma^{z}_j$ is a sum  over individual Pauli operators $\hat\sigma^{z}_j$  (each of them acts upon $j^{\rm th}$  two-level system in the ensemble). The uppering/lowering operators of a  collective "spin",    $\hat S^{\pm}=\sum\limits_{j=1}^N   \hat\sigma^{\pm}_j$,   are  also sums over respective  $\hat\sigma^{\pm}_j$. The light-matter coupling is encoded by two  last terms in (\ref{h}): the rotating-wave term with the coupling strength $g$   corresponds to the resonant interaction,     and  the counter-rotating term with $J$ corresponds to  the anti-resonant one.

 A rigorous field-theoretical description of the superradiant phase transition  in thermodynamic limit, $N\to \infty$,
 was proposed by  Popov and  Fedotov~\cite{popov1988functional}  in Matsubara formalism.
 The solution was obtained   in the rotating wave approximation (RWA), when anti-resonant terms are neglected, i.e.,   $J=0$. This case is also known as the Tavis-Cummings model (TCM).
 The phase transition is of $2^{\rm nd}$ order, it occurs if the    temperature is lower than a critical value $T<T_c$. The coupling constant must be higher than a critical value, $g>g_c$, otherwise, the system remains in normal phase for any temperature.  
 The   critical coupling $g_c=\sqrt{\omega\epsilon }$  does not depend on  $N $    due to $1/\sqrt N$ normalization in (\ref{h}); the   critical temperature is $T_c= \epsilon  \left(2 \ {\rm arctanh} \frac{ g_c^2}{g^2}\right)^{-1}$.

  Phase transition in RWA  was also studied in alternative situations. They include a regime of fixed excitations  density   and   finite chemical potential~\cite{eastham2001bose, eastham2006finite}, and   
  zero temperature regime   
 ~\cite{Pogosov_2017} where  Bethe ansatz technique was applied. 
  The study of fluctuational  normal-to-superradiant transition at finite temperatures and beyond the thermodynamic limit 
      was presented in Ref.~\cite{PhysRevA.99.063821}.

   According to a contemporary view on normal-to-superradiant quantum phase transition (QPT) in the symmetric Dicke model with $J=g$, it is characterized by   
		quantum chaotic dynamics~~\cite{PhysRevLett.90.044101, PhysRevE.67.066203} and dissipationless thermalization~\cite{PhysRevLett.108.073601}.
		 QPT is  of  $2^{\rm nd}$   order as   in RWA, however, the critical coupling      is  
$ g_c/2$.   The analysis of scaling behavior near QPT at finite-$N$ was provided in Ref.~\cite{refId0}. 
 Recently, an analysis of   quantum chaos in the symmetric Dicke model via  of out-of-time-ordered correlators attracted a great interest~\cite{PhysRevA.99.043602,Lewis-Swan2019,PhysRevLett.122.024101}. 
 
 We  also note that the superradiance is not a unique  QPT  in the 
 this model. Another one 
 is known as classical oscillator limit of $\omega=0$, where  finite-$N$ phase diagram of a ground state is  rather rich showing (non-)  and critical entanglement~\cite{PhysRevA.85.043821}.

\begin{figure}[htp]
	\center{\includegraphics[width=0.75\linewidth]{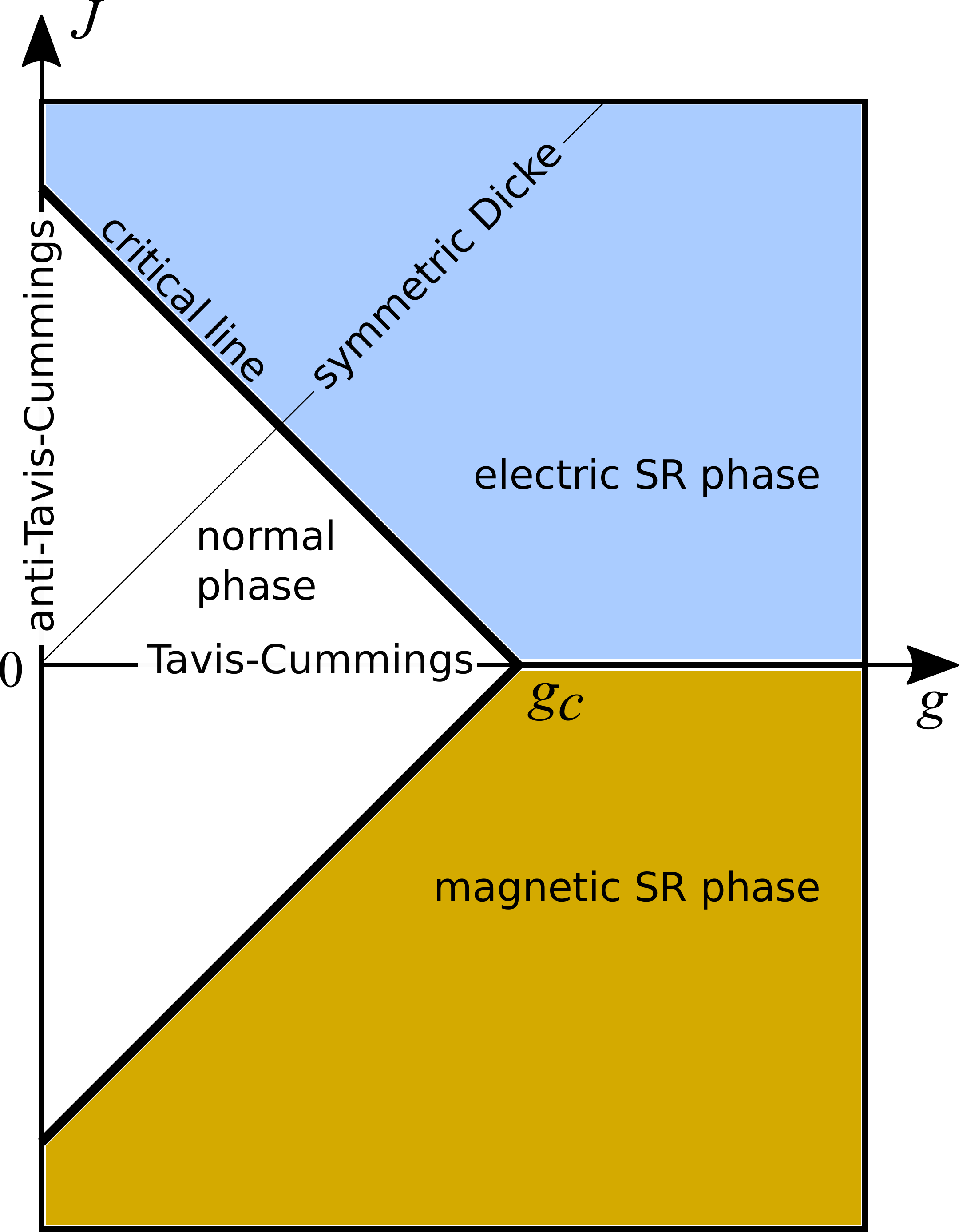}
	}
	\caption{Phase diagram of 
	GDM for thermodynamic limit of $N=\infty$. Horizontal (vertical) axis corresponds to TCM (anti-TCM); thin line $J=g$ corresponds to symmetric Dicke model. Critical lines, determined by the relation $g+|J|=g_c$, correspond to     transitions from  normal to "electric" ($J>0$) or "magnetic" ($J<0$) superradiant (SR) phases.  } \label{ph-diag-0}
\end{figure}

Generalizations of QPT on the case of $g\neq J$ were studied    in Refs.~\cite{ALCALDE20113385,Alcalde_2007,PRL_119_220601,PhysRevLett.112.173601}. 
Schematic phase diagram in thermodynamic limit is depicted in Fig.~\ref{ph-diag-0}.
       The superradiant phase exists in the domain  defined by   $ g + |J| >g_c$ which determines $2^{\rm nd}$ order phase  transition. 
         If  $g>0$  but $J$ may change its sign, then    $J=0$ is  the line of  $1^{\rm st}$ order phase  transition between the superradiant phases      of  "electric" 	($ J >0$) and "magnetic" ($  J  <0$) types~\cite{PhysRevLett.112.173601}.   The criticality and diagram of  "magnetic" and  "electric" phases 
         	  at finite $N$ were 
         	  analyzed
         	  in Ref.~\cite{PRL_119_220601}.

  In contrast to the previous studies of  QPT in generalized model, 
 we  study  the regime of 
 finite $T$ and  $N$     in the present paper.  Employing  the path integral  approach  of Ref.~\cite{PhysRevA.99.063821},  we go beyond    RWA to an asymmetric light-matter interaction  and explore   photon condensate fluctuations   near the normal-to-superradiant transition.     	Strictly speaking, we deal not with the phase transition in its conventional mean-field sense 
 but with a fluctuational transition of a finite width. 
As for any critical region, a natural question  on the corresponding fluctuational behavior 
  arises.  We show that the critical region   has rather complicated internal structure, where relative fluctuations of photon number and field squeezing 
  reveal different universal behaviors.

The paper is organized as follows. In Sec. \ref{Methodology} the problem formulation and path integral methodology are introduced.
In Sec.~\ref{Results} the results of the work are presented:  relative fluctuations,   Fano factor and squeezing parameters 
are  discussed    in~\ref{Universal_fluctuations},  
 minimal   temperature scales of our theory are found in~\ref{Minimal_temperature},    and   alternative approach based on Bethe ansatz  is introduced in~\ref{Bethe_ansatz}. 
In Sec.~\ref{Discussion} we discuss  our results and in Sec.\ref{summary} we  summarize.

   \section{Methodology}
   \label{Methodology}
    \subsection{Problem formulation}\label{Problem_formulation}
  Eigenfunctions of the model (\ref{h}) have  infinitely entangled structure    due to discrete $\mathbb{Z}_2$ symmetry when  $g$ and $J$ are simultaneously non-zero.  In this   case, there is only conservation of the parity of total excitations number.    The Hamiltonian  commutes  
        with the parity operator   $\hat \Pi=\exp\left[i \pi \hat M_+ \right]$ where  $\hat M_+=\hat a^\dagger\hat a + \frac{1 }{2}  \hat S^z$ is the operator of the total excitations number. However, there are two particular limits where the Hamiltonian possesses a continuous $U(1)$ symmetry   and becomes integrable. The first case is TCM,   realized when $J=0$ and $g\neq0$. Here, $\hat H$  conserves total excitations number, i.e. $\hat H $ and $\hat M_+$ commute.
    	 The second case is  anti-Tavis-Cummings  model (anti-TCM), realized when $g=0$ and $J\neq0$. This is nothing but opposite to RWA limit when  the only anti-resonant term  appears  in  (\ref{h}). The corresponding $\hat H$    conserves the excitation number difference defined through the operator $\hat M_-=\hat a^\dagger\hat a - \frac{1 }{2}\hat S^z$. Here,   $[\hat H, \hat M_-]=0$ and this   is another type  of  continuous $U(1)$ symmetry.      
    
    Interaction   parameters   are assumed to be non-negative  throughout the paper, $g\geq 0$ and $J\geq 0$, hence, we address  
the  superradiant phase  of   "electric" type   according to
    	Ref.~\cite{PhysRevLett.112.173601} (see Fig.~\ref{ph-diag-0}).

	In   superradiant phase a respective symmetry of $\hat H$,  $\mathbb{Z}_2$   or $U(1)$,  is broken and photons form a superradiant condensate.   In thermodynamic limit, the critical line of the phase transition is  $g+J=g_c$.  For $N\neq \infty$ the critical line is smeared into a fluctuational region of   finite width where average photon number changes smoothly.    We are focused on      equilibrium properties of photon condensate 
    into this critical region and 
 analyze relative fluctuations parameter 
 \begin{equation}
 	r   = \frac{\langle\!\langle (\hat a^\dagger \hat a)^2 \rangle\!\rangle_\beta}{ \langle    \hat a^\dagger \hat a \rangle_\beta^2}   \ , 
 	\label{r} 
 \end{equation}
     Fano factor 
 \begin{equation}
 	F   = \frac{\langle\!\langle (\hat a^\dagger \hat a)^2 \rangle\!\rangle_\beta}{ \langle    \hat a^\dagger \hat a \rangle_\beta}   \ , 
 	\label{F} 
 \end{equation}
and   squeezing parameters 
 \begin{equation}
 	\delta x=\frac{1}{2} \sqrt{ \langle\!\langle \hat x^2 \rangle\!\rangle_\beta} \ , \quad \delta p   =
 	\frac{1}{2} \sqrt{ \langle\!\langle  \hat p^2 \rangle\!\rangle_\beta}   \ .
  	\label{squee} 
 \end{equation}
The canonical coordinate $\hat x =(\hat a^\dagger +\hat a)/\sqrt2$ and   momentum $\hat p=i(\hat a^\dagger -\hat a)/\sqrt2$ correspond to electric and magnetic fields. 
  Here    
 $\langle \hat {\mathcal{O}} \rangle_\beta={\rm Tr} [ \hat    {\mathcal{O}} e^{-\beta \hat H}  ]/{\rm Tr} [   e^{-\beta \hat H}  ] $ denotes thermodynamical averaging, where $e^{-\beta \hat H}$ is equilibrium density matrix,     $\beta=1/T$,  and fluctuations $\langle\!\langle \hat {\mathcal{O}}\rangle\!\rangle_\beta=\langle  \hat {\mathcal{O}}^2 \rangle_\beta-\langle  \hat {\mathcal{O}}\rangle _\beta^2$.

\begin{figure}[htp]
	\center{\includegraphics[width=0.99\linewidth]{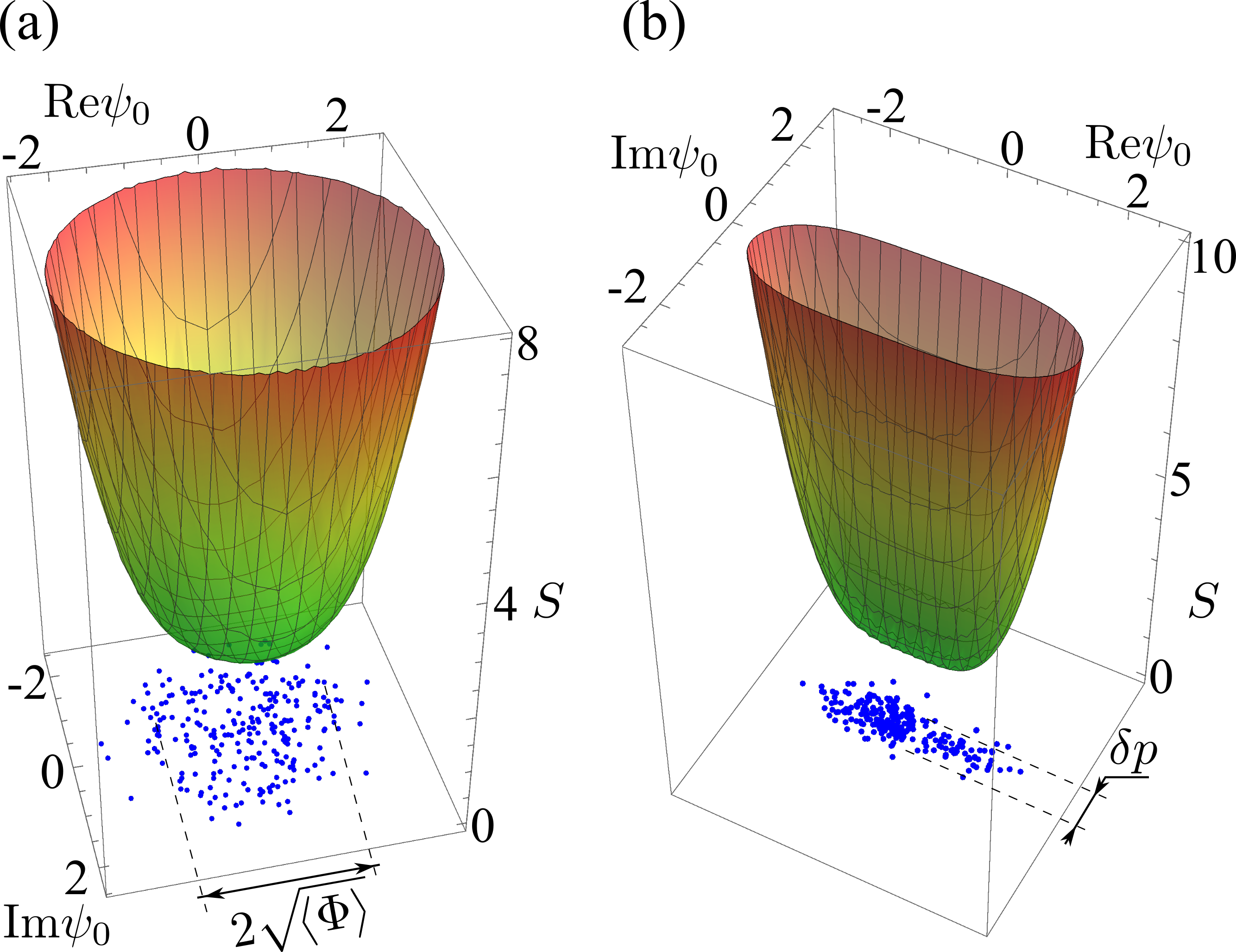}}
	\caption{ Effective potential $S$ for condensate mode $\psi_0$ at the critical point of the superradiant phase transition. [panel (a)]:  $U(1)$ case of TCM with Goldstone potential.  [panel (b)]: $\mathbb{Z}_2$ case of GDM with non-Goldstone potential.  250 blue dots in each panel correspond to  numerical simulation of a random realization of  $\psi_0$  with the respective $S$. The variance of the random $\psi_0$ in  TCM is given by the average photon number as  $2\sqrt{\langle\Phi\rangle}$, the squeezing in GDM is shown as $\delta p$; their expressions are given in (\ref{Q}) and (\ref{p_max}).
	Parameters of the simulation: $N=50$, temperature $T=\omega/10$,   $\epsilon=\omega$, $J=0$ and $g=g_c$ in (a), and $J=g=g_c/2$ in (b). }  \label{potential}
\end{figure}

In our approach, thermodynamic averages are calculated  by means of a path integral  and  Matsubara effective action. The action is formulated 
 for complex boson field $\psi_\tau$ defined on imaginary 
 time $\tau\in [0; \ \beta]$. 
This field and its conjugate, $\bar\psi_\tau$, correspond to operators $\hat a$ and $\hat a^\dagger$, respectively. 
As known from previous works~\cite{popov1988functional,eastham2001bose},   zero Matsubara mode $\psi_0$,     which is a complex variable,  parametrizes  superradiant   order parameter. It can be represented as  $\psi_0=\sqrt{\Phi}e^{i\varphi}$ where $\Phi$ and $\varphi$ are real variables in a path integral. They have a transparent meaning:  the magnitude $\Phi$  is  related to a    photon number in the condensate, and $\varphi$ is the order parameter complex phase. 
The zero-frequency mode is highlighted  relative to others  because it  corresponds   to spontaneously  emergent  non-zero average of the photon field.   The Goldstone effective potential $S[\bar\psi_0;\psi]$ for $U(1)$ case is shown in Fig.~\ref{potential}~(a); blue dots and the variance, $2\sqrt{\langle\Phi\rangle}$, corresponds to numerical simulation of random $\psi_0$ distributed with the probability density $\propto e^{-S[\bar\psi_0;\psi]}$.

A  consequence of $\mathbb{Z}_2$ symmetry   is that fluctuations of $\Phi$ and $\varphi$
are governed by a non-Goldstone effective potential as shown in Fig.~\ref{potential} (b). 
Hence, relative fluctuations 
in the critical region are determined not only by fluctuations of $\Phi$, as that in $U(1)$ TCM~\cite{PhysRevA.99.063821}, but also by  fluctuations of condensate's phase which gives a non-trivial contribution.    
For instance, in $U(1)$ case the squeezing is absent, while it appears in the generalized model under the consideration.  The effect of squeezing and respective parameter $\delta p$ are illustrated in Fig.~\ref{potential}~(b) for  the random distribution of $\psi_0$.

\begin{figure}[htp]
	\center{\includegraphics[width=0.95\linewidth]{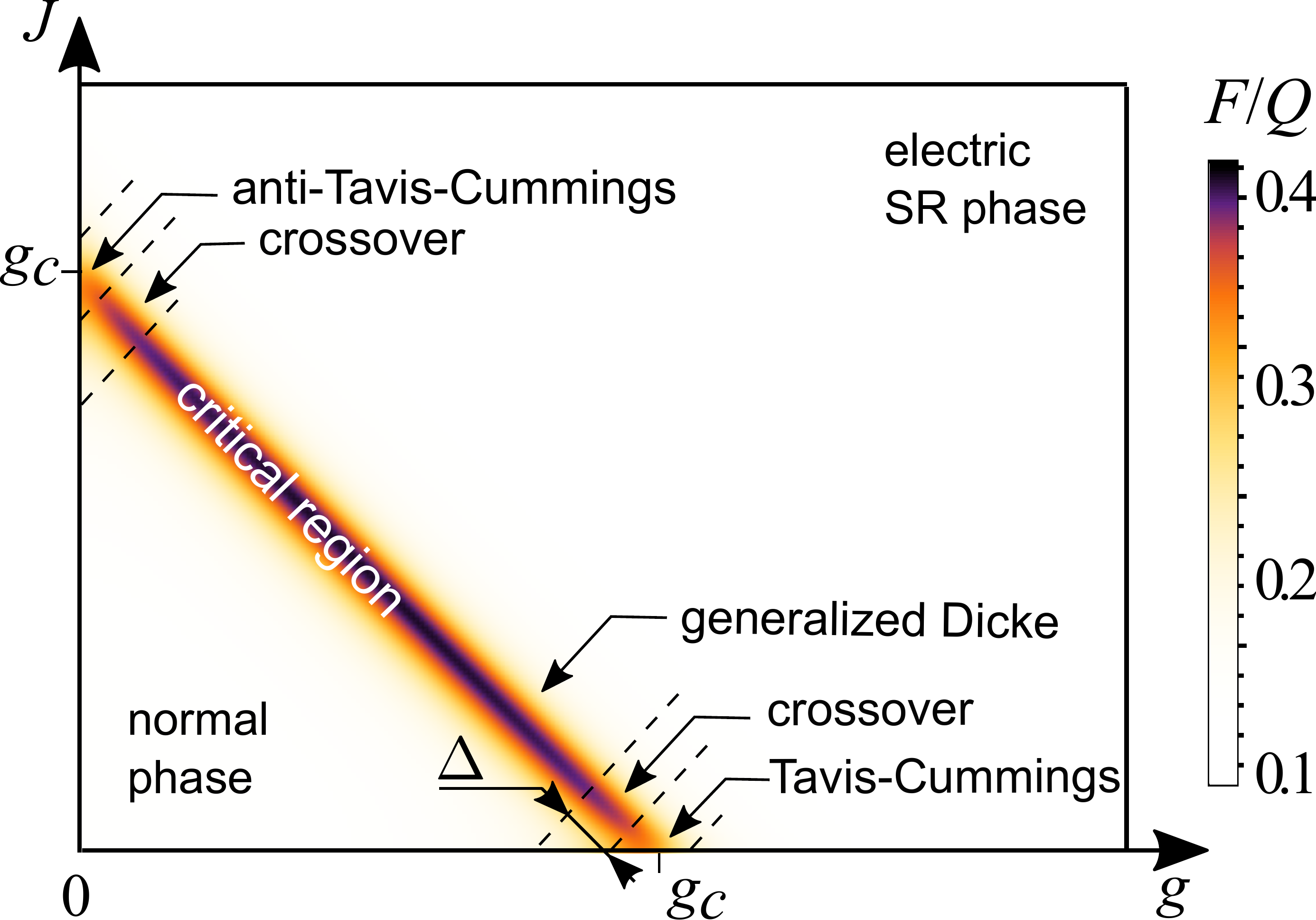}}
	\caption{Phase diagram of GDM at finite $N=50$, temperature $T=\omega/10$, and  $\epsilon=\omega$.  
	The Fano factor $F(g,J)$  normalized by $\mathcal{Q}=\sqrt{NT\epsilon/\omega^2}$ is plotted. 
 Normal phase  and electric superradiant phase (white regions)   are mediated by the critical region  of the  width $\Delta=\sqrt{\omega T/N}$ (colored area near the line $g+J=g_c$).   TCM and anti-TCM sectors, where the Fano factor takes universal ratio $F_{\rm TC}/\mathcal{Q}= \frac{\sqrt\pi}{2}-\frac{1}{\sqrt\pi}\approx 0.32204   $,
  and crossovers sectors are also determined by $\Delta$.  In GDM sector, which covers the major part of the critical region,  universal ratio is $F_{\rm GD}/\mathcal{Q}= \frac{   \Gamma  \left(5/4\right)   }{ \Gamma  \left(3/4\right)}  - \frac{  \Gamma  \left(3/4\right)  }{4\Gamma  \left(5/4\right) }   \approx 0.40168  $.  } \label{fig-diagram-F}
\end{figure}

According to Ref.~\cite{PhysRevA.99.063821},      TCM has  universal value of  $r=\frac{\pi}{2}-1$ at the critical region; its  width is determined by the scale $\Delta=\sqrt{\omega T/N}$. The Fano factor was shown to have a peak with the value much greater than unity, $F\gg 1$, which indicates for strongly positive correlations between photons at the phase transition. In normal and superradiant phases, however, $F<1$ and correlations are negative   (anti-bunching effect).  In this work   we analyze how $r$ and $F$ change if anti-resonant $J$ appears in the model.   Fig.~\ref{fig-diagram-F}   shows  the     phase diagram of the model (\ref{h}) with anti-resonant terms and finite $N$ and $T$.  The critical region of the interest   corresponds to the colored area (here  $F>1$ and the   width is also determined by  $\Delta$). The effective theory presented  allows  to analyze a behavior inside the critical region and describe fluctuations in TCM, anti-TCM and GDM sectors, as well as in crossovers between  them.

\subsection{Total   action 
}\label{Effective_action}
 In this part we introduce total   action 
 of a hybrid system at equilibrium described by the Hamiltonian (\ref{h}). 
 As indicated above,   when we formulate path integral technique, the photon mode is represented via boson complex field $\psi$.    However, Pauli operators    can be    parametrized in different ways in path integrals. 
 This can be Holstein-Primakoff bosonization  which provides an  exact diagonalization of the symmetric Dicke model  in  thermodynamic limit~\cite{PhysRevE.67.066203}.
Alternatively to the bosonization, there are several fermion representations.   For instance,      Pauli operators 
can be  parametrized via bilinear forms of semi-fermion fields~\cite{popov1988functional}.   These are Grassmann  fields with   unconventional boundary conditions on the imaginary time axis. 
 Another example   is a combination of conventional fermions  where an auxiliary boson field    is introduced in order to preserve the correct dimensionality of the  Hilbert space~\cite{eastham2001bose}.

In our approach  we choose    Majorana fermion representation of Pauli operators~\cite{martin1959generalized, tsvelik2007quantum, SCHAD2015401}. As shown in Ref.~\cite{PhysRevA.99.063821} for 
 TCM, this method is rather convenient for analysis of a fluctuational behavior near the phase transition. 

The Majorana representation of Pauli operator for $j^{\rm th}$ qubit  is defined through the bilinear form of 
a conventional (complex) fermion operator $\hat c_j\neq \hat c^\dagger_j$ and Majorana one $\hat d_j=\hat d_j^\dagger$ 
\begin{equation}
\hat\sigma^+_j=\sqrt{2}\hat c^\dagger_j \hat d_j, \quad \hat \sigma^-_j=\sqrt{2}\hat d_j\hat c_j \ .  \label{fermionization}
\end{equation}
The Majorana mode has zero energy with the average  $\langle \hat d^2_j\rangle_\beta=1/2$, while     complex fermion   has energy $\epsilon_j$.  The partition function as a path integral  reads as
\begin{equation}
	Z=
	\int \mathcal{D}[\Psi, \mathcal{C}  ]
	\exp(- S_{\rm tot}[\Psi, \mathcal{C}])  \ , \label{Z-tot}
\end{equation}
where complex boson variables are collected in the vector \begin{equation}\Psi_\tau^T(\tau)=[\bar\psi(\tau),\psi(\tau)]\end{equation}
 and independent Grassmann variables, which parametrize  fermion operators   $\hat c_j^\dagger$, $\hat c_j$, and $\hat d_j$, are collected in  the vector  \begin{equation} \mathcal{C}^T(\tau) =\{  \bar c_j(\tau) ,c_j(\tau),d_j(\tau) \}_{j=1}^N \ . \end{equation} 
Total Matsubara action 
is
 	 \begin{multline} S_{\rm tot} [\Psi, \mathcal{C} ]=  S_{\rm ph}[\Psi ] +  S_{\rm \sigma} [  \mathcal{C} ] + \\ + S_{\rm int}[\Psi, \mathcal{C}  ]+\ln Z_{\rm ph}Z_{\rm \sigma}  \ . \label{S} \end{multline}

The first 
term here  is the  free  photon mode's action 
\begin{equation}
	S_{\rm ph}[\Psi]  =\beta\sum\limits_{n }\bar\psi_n (-G_{{\rm ph}; n}^{-1} ) \psi_n \ , 
	\label{Sph-1}
\end{equation}
where the respective bosonic Green function is
\begin{equation}
 \quad G_{{\rm ph};  n}^{-1}={\rm i}2\pi n T -\omega \   .
	\label{Gph-1}
\end{equation}
The Matsubara modes $\psi_n$ with  $n\in \mathbb{Z}$ here are given by a discrete Fourier transformation on imaginary time interval      \begin{eqnarray}\psi_n=T\int\limits_0^\beta \psi(\tau) e^{i 2\pi n T \tau } d\tau \ , \nonumber  \\  \bar\psi_n=T\int\limits_0^\beta \bar\psi(\tau) e^{-i 2\pi n T \tau } d\tau \ . \end{eqnarray}  
 They correspond    to bosonic   frequencies $2\pi n T$. 
The same transformation applies  for fermion modes with odd   frequencies $(2\pi n +\pi)T$.

The second term in (\ref{S}) is the Majorana representation of  two-level systems'   action 
\begin{equation}
	S_{\rm \sigma}[ \mathcal{C} ]  
	=\frac{1}{2}\sum\limits_{j=1}^N\sum\limits_{n } \mathcal{C}_{j; -n}^T 
	(-\mathbf{G}_{n}^{-1} )\mathcal{C}_{j; n}
	\ , 	\label{Sq} 
\end{equation} 
where the inverse fermion Green function  has the following matrix structure
  \begin{multline}	 \mathbf{G}_{n}^{-1}{=}  
\begin{bmatrix} 0  & i(2 n{+}1)\pi T{-}  \epsilon & 0 \\ \\
	i(2 n{+}1)\pi T {+}\epsilon & 	0   & 0 \\ \\
	0  & 0 & i(2 n{+}1)\pi T 
\end{bmatrix} 
\nonumber
\end{multline}
 and acts on the vectors composed of Matsubara modes $\mathcal{C}_{j; n}=[\bar c_{j; -n} ; c_{j; n} ;  d_{j; n}]^T$. 
 
 The third term in (\ref{S}) is the interaction 
 \begin{equation}
	S_{\rm int}[\Psi, \mathcal{C} ] = \frac{1}{\sqrt{2} }\sum\limits_{j=1}^N\sum\limits_{m,k } \mathcal{C}_{j; -m}^T
	\mathbf{V}_{m-k}
	\mathcal{C}_{j; k} 	  	\label{Sint}
\end{equation}
  represented via the  matrix   $\mathbf{V}$   involving   complex boson fields as its elements:
 \begin{equation}
 	\mathbf{V}_{n}
	= 
	\begin{bmatrix} 0  & 0 & g\bar\psi_{-n}{+}J\psi_n \\ \\
	0 & 	0   &   -(g\psi_n {+} J \bar\psi_{-n}) \\ \\
	-(g\bar\psi_{-n} {+} J\psi_n)  & g\psi_n {+} J \bar\psi_{-n} & 0
	\end{bmatrix} \ .
\nonumber
\end{equation}
 Note that $m$ and $k$ indices in (\ref{Sint}) stand for fermionic frequencies, $(2\pi m+\pi)T$ and $(2\pi k+\pi)T$, while their difference in $\mathbf{V}_{m-k}$ stands for bosonic one $2\pi(m-k)T$.
 
 The last term  in (\ref{S}) provides the  unity normalization of $Z$ for a non-interacting limit  $g=J= 0$. The partition functions of free photon mode, $  Z_{\rm ph}=\prod\limits_n (-G_{{\rm ph};n})$, and isolated $N$ two-level systems, $Z_{\rm \sigma}=\prod\limits_n \big( {\rm Det } \ (-{\mathbf{G}}_{n} )\big)^{-N/2}$,  are given by infinite products over Matsubara modes, as follows from Gaussian integration rules. 
 They read as  \begin{multline}
 \int\!\! D[\bar\Psi,\Psi] e^{-\bar \Psi  A \Psi } = \\ = \int\prod\limits_n\frac{d\bar \psi_n d\psi_n}{\pi}\exp\left[-\sum\limits_{n,m}\bar \psi_n A_{n,m} \psi_m \right]=
 \frac{1}{{\rm Det} A }\ , 
 \end{multline}
 for complex  variables (the matrix $A$ has non-negative eigenvalues). 
 For non-independent Grassmann variables with an anti-symmetric matrix $\mathcal{A}$ we have
 \begin{multline}
\int D[\mathcal{C}] e^{-\frac{1}{2}\mathcal{C}^T \!\mathcal{A}  \mathcal{C} } = \\ = \!\int\prod\limits_{j;n} d\bar c_{j;n} dc_{j;n} dd_{j;n} \exp\!\left[-\frac{1}{2}\! \sum\limits_{j;n,m}\!\mathcal{C}_{j;n} \mathcal{A}_{j;n,m} \mathcal{C}_{j;m}\right]\!=\\
=\sqrt{{\rm Det} \  \mathcal{A} }  \ . \label{Gauss-int-fermion}
 \end{multline}

 \subsection{Effective functional  for photon mode fluctuations}
In this part an effective action for equilibrium photon mode is derived. We start from   Gaussian  integration over $\mathcal{C}$-fields  with the use of the identity (\ref{Gauss-int-fermion}) equivalent to taking a  trace over the Hilbert space of two-level systems.
 Applying the  identity 	\begin{equation}
 	\ln {\rm Det}  \ \mathcal{A}={\rm Tr} \ln  \mathcal{A}  \ ,
 \end{equation}
we arrive at the effective action in a general form $S_{\rm eff}[\Psi]=S_{\rm ph}[\Psi] +\ln Z_{\rm ph}Z_{\rm \sigma}  -  \frac{1}{2}N
	 {\rm Tr} \ln (-\mathbf{G}^{-1}  + \mathbf{V})$, which, after  a standard resummation of the logarithm, becomes 
 \begin{multline}
	 S_{\rm eff}[\Psi]=
	 S_{\rm ph}[\Psi] +\ln Z_{\rm ph}\sqrt{Z_{\rm \sigma} } - \\ -\frac{1}{4}N
	 {\rm Tr} \ln \left[-\delta_{n,m}\mathbf{G}^{-1}_m  +\sum\limits_l \mathbf{V}_{n-l}\mathbf{G}_l\mathbf{V}_{l-m}\right] \ . \label{s_eff_1}
\end{multline}
To obtain the effective functional  from (\ref{s_eff_1}) describing the superradiant phase transition and fluctuations above the photon condensate, we  
separate zero mode from the others  in the self-energy matrix as
 \begin{multline}
\sum\limits_l \mathbf{V}_{n-l}\mathbf{G}_{j;l}\mathbf{V}_{l-m} = \\ =\delta_{n,m}\mathbf{V}_{0}\mathbf{G}_m\mathbf{V}_{0} + 
\sum\limits_{l\neq n,  m} \mathbf{V}_{n-l}\mathbf{G}_{j;l}\mathbf{V}_{l-m} + \\
+(1-\delta_{n,m})(\mathbf{V}_{n-m}\mathbf{G}_m\mathbf{V}_{0}+\mathbf{V}_{0}\mathbf{G}_n\mathbf{V}_{n-m})  \ .  \label{GVG}
\end{multline}
The first term  depends  on zero mode $\psi_0$ only, while the second term   is a non-diagonal matrix that is determined by non-zero   modes $\psi_{n\neq 0}$ which  describe quasiparticle fluctuations above the condensate.  The third term     is a product of  zero and non-zero Matsubara modes; it  cancels out in further calculations.   

At this step we introduce new fermion Green function which absorbs the diagonal part as 
  \begin{equation}
 	\mathcal{G}_{j; m}[\bar\psi_{0},\psi_{0}]= \left[\mathbf{G}^{-1}_{j;m} -   \mathbf{V}_0\mathbf{G}_{j;m} \mathbf{V}_0 \right]^{-1}\ 
 \end{equation}
 and expand the logarithm in $S_{\rm eff}[\Psi]$ by a first order  in the non-diagonal part of  $\mathbf{V}\mathbf{G}\mathbf{V}$:
 \begin{multline}
\ln  \sqrt{Z_{\rm \sigma} } -  \frac{1}{4}N{\rm Tr} \ln \left[-\delta_{n,m}\mathbf{G}^{-1}_m  +\sum\limits_l \mathbf{V}_{n-l}\mathbf{G}_l\mathbf{V}_{l-m}\right] \approx \\
\approx -  \frac{1}{4}{\rm Tr} \ln\left( \mathbf{G}_{j;m}\mathcal{G}^{-1}_{j;m}[\bar\psi_{0},\psi_{0}]\right)+  \\ +\frac{1}{4}N{\rm Tr} \left[ \mathcal{G}_{j; n}[\bar\psi_{0},\psi_{0}]  \sum\limits_{l\neq n,m} \mathbf{V}_{n-l}\mathbf{G}_{j;l}\mathbf{V}_{l-m}\right] \ . \label{expansion}
\end{multline}
Here we use   the assumption  that quasiparticle fluctuations $\psi_{n\neq 0}$ are sufficiently small. 

The effective action takes the following form after the expansion (\ref{expansion}):
 \begin{equation}
	S_{\rm eff}[\Psi]\approx  S [\bar\psi_0; \psi_0]+S_{\rm fl}[\bar\psi_{n\neq 0}; \psi_{n\neq 0}]+\ln Z_{\rm ph} \ .   
	\label{s_eff}
\end{equation}
The first term is the functional for superradiant condensate
 \begin{multline}
S[\bar\psi_0; \psi_0] = \\ \!\!\! = -\beta G_{{\rm ph}; 0}^{-1} |\psi_0|^2 - 
N \ln \frac{\cosh\frac{\sqrt{\epsilon^2+4|g\psi_0+J\bar\psi_0|^2/N 
	}}{2T}}{\cosh\frac{ \epsilon} {2T}}
\ .   \label{s-0}
\end{multline}
The logarithmic term here is a result of a calculation of ${\rm Tr} \ln\left( \mathbf{G}\mathcal{G}^{-1}[\bar\psi_{0},\psi_{0}]\right)$ in the second line of (\ref{expansion}), where the trace    is reduced to a calculation of the infinite product over fermion Matsubara modes. The functional (\ref{s-0}) is of a non-Goldstone type due to a dependence on a complex phase of $\psi_0$.

The second term in (\ref{s_eff})   is responsible for  Gaussian fluctuations  above the condensate 
\begin{equation}S_{\rm fl}[\bar\psi_{n\neq 0}; \psi_{n\neq 0}]=
	\frac{\beta}{2}\sum\limits_{n\neq 0}   \Psi_{-n}^T (-G_{{\rm fl}; n}^{-1}[\bar\psi_0; \psi_0])\Psi_n  
	\ ,   \label{S-fl-0}
\end{equation}
where $\ \Psi_n^T=[\psi_n, \bar\psi_{-n}]$. The  inverse Green function matrix  $G_{{\rm fl}; n}^{-1}[\bar\psi_0; \psi_0]$ involves self-energy   given by the last term in (\ref{expansion}). Formally, this self-energy does depend on the zero mode variable. However, such a  dependence provides small by $1/N$ corrections   when the system is near the critical region. Consequently, $\psi_0$-dependence can be neglected  and we suggest $G_{{\rm fl}; n}^{-1}= G_{{\rm fl}; n}^{-1}[\bar\psi_0{=}\psi_0{=}0]$ where
\begin{multline} 
	G_{{\rm fl}; n}^{-1} = \\ \!\!\!	\begin{bmatrix}
		-gJ( \Sigma_n+ \Sigma_{-n}) &  G_{{\rm ph}; n}^{-1} {-} (g^2 \Sigma_{-n} {+} J^2\Sigma_{n}) \\  \\
		G_{{\rm ph}; n}^{-1} {-} (g^2 \Sigma_n {+} J^2\Sigma_{-n})  &  -gJ( \Sigma_n{+} \Sigma_{-n}) 
	\end{bmatrix} \!. 
	\!\!\!\!
\end{multline}
Self-energies are  parametrized by
\begin{equation} \Sigma_n=\frac{\tanh\frac{\epsilon}{2T}}{2i \pi n T -\epsilon} \ 
\end{equation}
which coincides with a self-energy in RWA.  

 Note that  the condensate functional $S[\bar\psi_0; \psi_0] $ is symmetric under the interchange of $g$ and $J$, however, the model (\ref{h}) does not have this symmetry for $\epsilon\neq 0$. As it should be, this asymmetry is recovered in the total action $S_{\rm eff}[\Psi]$ which involves  excitations above the condensate encoded by non-zero modes. It can be seen from  $S_{\rm fl}[\bar\psi_{n\neq 0}; \psi_{n\neq 0}]$, where $G_{{\rm fl}; n}^{-1}$   is not symmetric under the interchange of $g$ and $J$. Namely, the asymmetry  follows from  $\Sigma_n\neq \Sigma_{-n}$ for any $\epsilon\neq 0$. 
 
 \subsection{Effective action for  condensate magnitude}
 As long as we address to the  critical region near the superradiant transition,   the leading    contribution  to fluctuational behavior comes
 from the photon condensate.
 Hence, calculations of    thermodynamical average is reduced to path integral with the only one complex variable $\psi_0$.  As mentioned before, we parametrize it as
 \begin{equation}
\psi_0=\sqrt{\Phi}e^{i\varphi} 
 \end{equation}
  where  $\Phi=|\psi_0|^2$ is the  magnitude of  superradiant order parameter  and  $\varphi$ is its phase.  Both  $\Phi$ and $\varphi$ are quantum variables  fluctuating in 
 the potential  
\begin{multline}
		  S [\Phi, \varphi]=  \frac{\omega}{T}\Phi + N \ln \cosh\frac{ \epsilon  } {2T}
		-   \\   -  N \ln \cosh\left[ \frac{\epsilon}{2 T}\sqrt{1+\frac{4 \Phi }{N \epsilon^2 } (g^2{+}J^2{+}2g J \cos2\varphi ) 
		} \right]   
		 \! .  \! \!   \label{s-zm}
	\end{multline}

We study hereafter a behavior at  temperatures, $T\ll\epsilon$, where the  condensate functional (\ref{s-zm})  is reduced to the following form
\begin{multline}
	S [\phi, \varphi]= \phi - \frac{1}{2\gamma} \left(\sqrt{1+4\gamma \eta[\varphi]\phi}-1 \right) \ , \\ \eta[\varphi]  = \eta_0-2\eta_1\sin^2\varphi \ , \label{s-zm-0}
\end{multline}
where $\phi  =\beta\omega\Phi $ is  rescaled order parameter   and $\eta[\varphi] $ determines phase dependence. 
The dimensionless interaction parameters in (\ref{s-zm-0}) read
 \begin{eqnarray}
	&	\eta_0 &=  \frac{(g +J)^2}{\epsilon \omega}  \ , \label{eta0} \\
	&	\eta_1 &=  \frac{2gJ}{\epsilon \omega}  \ , \label{eta1}   
	\end{eqnarray}
	and rescaled temperature,
	 \begin{eqnarray}
	&	\gamma &=  \frac{T}{N \epsilon }\ , \label{gamma}
\end{eqnarray}
is small parameter of our theory, $\gamma\ll 1$.
 A remarkable property of the action (\ref{s-zm-0}) is that    $\gamma $ appears twice  as a denominator in $\frac{1}{2\gamma}$ and as a prefactor into the square root term.     This property allows to extract the relevant part of  the action $S [\phi, \varphi]$ within three steps.

  The first step is an   expansion of   the square root   by small $\gamma$ up to the second order. Here we assume that phase and magnitude are such that the condition $\gamma \eta[\varphi]\phi\ll 1$ is fulfilled.  Note, the presence of overall $\frac{1}{2\gamma}$ prefactor in (\ref{s-zm-0}) reduces the order of $\gamma$ in this expansion and the action at this point  reads as $S [\phi, \varphi]=(1-\eta[\varphi])\phi+ \gamma\eta^2[\varphi]\phi^2$. In particular, $\gamma$ cancels out in   front of the linear in $\phi$ term $(1-\eta[\varphi])\phi$; this fact is crucial for further analysis.   At the phase transition, the relevant values of phase are such that $\eta[\varphi] \sim 1$; hence  path integrals over $\phi$ converge in the domain $0 <\phi\lesssim \gamma^{-1/2}  
$.  This means that the initial condition $\gamma \eta[\varphi]\phi\ll 1$ is always satisfied and $\gamma$-expansion  is strict.

The second step is to neglect     $\varphi$-dependence    in the quadratic term $\gamma\eta^2[\varphi]\phi^2 $ and replace it by $ \gamma \eta_0^2 \phi^2$. Here the phase is replaced by its value at the functional minima ($\varphi=0$ and $ \varphi=\pi$) where $\sin\varphi=0$. The result of the above two step reads: 
\begin{equation}
	S [\phi, \varphi]=(1-\eta_0+2\eta_1\sin^2\varphi)\phi+\gamma\eta_0^2\phi^2    \ . \label{s-zm-exp}
\end{equation}

    The third step of our  derivation is the   integrating out   the phase $\varphi$ from the action (\ref{s-zm-exp}). This is performed  
exactly with the use of the identity \begin{equation}\int\limits_0^{2\pi}e^{-2z\sin^2\varphi} d\varphi=2\pi e^{-z}I_0(z) \ ,
\end{equation}
	where 
	$z=\eta_1\phi$ and $I_0(z)$ is modified Bessel function of zero order.  Ascending the result of integration into the exponent, we arrive at one of the central results of this work,  the  action for photon condensate magnitude: 
\begin{equation}
	S_\phi = (1-\eta_0+\eta_1) \phi+\gamma \eta_0^2\phi^2 -  \ln I_0(\eta_1\phi) \ .  \label{s-phi}
\end{equation}
The modified Bessel function logarithm,  $\ln I_0(\eta_1\phi)$, describes the dissipative dynamics of $\phi$ due to a coupling between the fluctuations of  the density of photon condensate and its phase.

 It is important  that the small parameter $\gamma$ does not enter into the dissipative term (\ref{s-phi}). Instead, $\gamma$ appears in quadratic term only and provides a width of Gaussian tails $\phi\sim 1/\sqrt\gamma$ in the partition function exponent.   It means that a    character    argument of the Bessel function is estimated as $z=\eta_1/\sqrt\gamma$ in this case. It  can be both  small or large   compared to unity ($z\ll1$ and  $z\gg 1$), depending on the interaction   parameters. Consequently, different asymptotic   expansions can be   applied for $\ln I_0(z)$.  
To the best of our knowledge,    properties of this action has not yet been studied     in GDM context and  our work is devoted to this issue.

\subsection{Expressions for $r$, $F$, and squeezing parameters 
 } \label{Main_definitions}
Before we proceed with asymptotic expansions of the dissipative term, let us make a step back to definitions for   $r$, $F$, and squeezing parameters.
According to the path integral approach indicated above, thermodynamical average of a certain operator $\mathcal{F}[\hat a, \hat a^\dagger] $ is represented via  integrals over the condensate variables $\Phi$ and $\varphi$. Hereafter, this  double integral is denoted as $\langle  \ \rangle$-brackets  without a subscript, and thermodynamical average then becomes: 
  \begin{multline}
  	\langle \mathcal{F}[\hat a, \hat a^\dagger] \rangle_\beta = \langle \tilde{\mathcal{F}}[\Phi,\varphi] \rangle  \ ,  \\ \langle \tilde{\mathcal{F}}[\Phi,\varphi] \rangle \equiv  Z_0^{-1}\int\limits_0^\infty\! d\Phi\int\limits_0^{2\pi}\!  d\varphi \  \tilde{\mathcal{F}}[\Phi,\varphi] e^{ -S [\Phi, \varphi]} \ . 
  	\label{p-int-def}
  \end{multline} 
Here $\mathcal{F}$ transforms into $\tilde{\mathcal{F}}$  under the parametrization  of  $\psi_0$ through $\Phi$ and $\varphi$, and $Z_0$ is normalization factor  providing $\langle 1 \rangle =1$.
The  same definition (\ref{p-int-def}) applies for thermodynamic   fluctuations: $
\langle\!\langle \mathcal{F} \rangle\!\rangle_\beta = \langle\!\langle  \tilde{\mathcal{F}}  \rangle\!\rangle  
$ with $\langle\!\langle  \tilde{\mathcal{F}}  \rangle\!\rangle\equiv\langle  \tilde{\mathcal{F}} ^2 \rangle - \langle  \tilde{\mathcal{F}}  \rangle^2$.
Namely, the  average  and fluctuations of   photon numbers,  i.e., their first and second cumulants, have the following   forms:
\begin{equation}
	\langle \hat a^\dagger \hat a \rangle_\beta = \langle \Phi \rangle \ , \quad \langle\!\langle (\hat a^\dagger \hat a)^2  \rangle\!\rangle_\beta = \langle\!\langle \Phi^2  \rangle\!\rangle \ . \label{aa}
\end{equation}
From technical point of view we use  the effective functional for rescaled order parameter (\ref{s-phi}) in calculation the photon number moments:
 \begin{equation}
	\langle  \Phi^k  \rangle  =  \frac{ \int\limits_0^\infty \phi^k  e^{-S_\phi}   d\phi }{   (\beta\omega)^k \int\limits_0^\infty    e^{-S_\phi}  d\phi   }  \ .  \label{cumulant}
\end{equation}
 The representation (\ref{cumulant}) is used in calculation of  $r$ and $F$ parameters.

The electric and magnetic fields operators  in $\Phi$-$\varphi$ representation become
\begin{equation}
	\frac{\hat a^\dagger +\hat a}{\sqrt2} \to  \sqrt{2\Phi}\cos\varphi  \label{el-field}
\end{equation}
and 
\begin{equation}
	i\frac{\hat a^\dagger -\hat a}{\sqrt2} \to  \sqrt{2\Phi}\sin\varphi \ , \label{m-field}
	\end{equation}
respectively. 
First cumulants of  these fields equal zero, $\langle \sqrt\Phi \cos\varphi\rangle=\langle \sqrt\Phi \sin\varphi\rangle=0$, due to $\pi$-periodicity of $S [\Phi, \varphi]$ functional. As a consequence, second cumulants of $\delta x$ and $\delta p$  coincide with the respective second moments.  The expressions for squeezing parameters (\ref{squee}) in $\Phi$-$\varphi$ representation become  averages of the fields quadratures (\ref{el-field}, \ref{m-field}):
\begin{equation}
	\delta x =\sqrt{\langle \Phi \cos^2\varphi\rangle} \ , \delta p =\sqrt{\langle \Phi \sin^2\varphi\rangle} \ . \label{squee-1}
\end{equation} Calculation of the squeezing parameters (\ref{squee-1}) is based on $S [\phi, \varphi]$ from (\ref{s-zm-exp}). The integration over $\varphi  $ is performed analytically  and expressions for $\delta x$ and $\delta p $ are represented as $\phi$-integrals:
 \begin{eqnarray}
 	  \delta x    =  \left[ \frac{ \int\limits_0^\infty \phi\big(1+ \frac{I_1(\eta_1\phi)}{I_0(\eta_1\phi)}\big)  e^{-S_\phi}   d\phi }{ 2 \beta\omega \int\limits_0^\infty    e^{-S_\phi}  d\phi   } \right]^{1/2}  \ , \nonumber \\
	  \delta p   =  \left[ \frac{ \int\limits_0^\infty \phi\big(1- \frac{I_1(\eta_1\phi)}{I_0(\eta_1\phi)}\big)  e^{-S_\phi}   d\phi }{ 2 \beta\omega \int\limits_0^\infty    e^{-S_\phi}  d\phi   } \right]^{1/2} \ .  \label{sq-x-p}
 \end{eqnarray}

It should be noted, the higher-order corrections  due to non-zero modes $\psi_{n\neq 0}$  are neglected here by small parameter $T^*/T$, where $T^*$ is a minimal temperature scale where our theory can be applied, i.e. when the first order expansion in (\ref{s_eff}) is correct. 
 We discuss this issue in more detail in  Subsection~\ref{Minimal_temperature}.

\section{Results} \label{Results}
\subsection{Universal fluctuations at the critical region
} \label{Universal_fluctuations}

  Depending on  a value of  $\eta_1$,  there are different universal behaviors of fluctuations at the phase transition. They are dictated by the dissipative term $\ln I_0(\eta_1\phi)$ in the action $S_\phi$.
 As mentioned above,  the relevant values of $\phi $, where the path integrals (\ref{cumulant}, \ref{sq-x-p})  converge, are determined by the scale of $\gamma^{-1/2}$. 
Introducing  a parameter $z=\eta_1\gamma^{-1/2}$ as a typical scale of the dissipative term argument, there are different   asymptotics of $\ln I_0(\eta_1\phi)$. We address the important limits of small, $z\ll 1$, and large, $z\gg 1$,  parameters. 

As shown below, the limit of $z\ll1$  corresponds to TCM and anti-TCM regimes  (see  Fig.~\ref{fig-diagram-F}) where the anti-resonant interaction is effectively suppressed by thermal fluctuations. Oppositely,   $z\gg 1$, is related to GDM where anti-resonant interaction becomes relevant. Also, an  intermediate regime of  $z\sim 1$ is  analysed; it corresponds to a crossover between   universal behaviors of TCM and GDM located inside the critical region.

\subsubsection{(Anti-) and Tavis-Cummings regimes}\label{anti_TC}

Here we address to $z\ll 1$ asymptotic of (\ref{s-phi}), in other words, the limit of small anti-resonant interaction $\eta_1 \ll \sqrt\gamma\ll 1 
 $. This range determines the interaction parameters near the point $\eta_1=0$, which, according to (\ref{eta1}), corresponds to   $U(1)$ limits of (anti-) and TCM.  Note that these models reveal identical structures of condensate functionals. As mentioned above, this fact is a consequence of  the consideration limited by  zero mode only.

The small argument expansion for the dissipative term up to fourth order is \begin{equation}\ln I_0(z\ll 1)\approx \frac{1}{4}z^2 - \frac{1}{64}z^4 \ . \label{LogI0-small}
\end{equation}
	 Hence, one arrives at the following expression:
\begin{equation}
	S_\phi = (1-\eta_0+\eta_1) \phi+\left(\gamma \eta_0-\frac{\eta_1^2}{4}\right)\phi^2  +\frac{\eta_1^4}{64}\phi^4 \ .  \label{s-phi-small}
\end{equation}
It is important to note that contributions of third and fourth orders by $\phi$ also appear in the square root expansion of (\ref{s-zm-0}), however they are small by $\gamma$ compared to those given by $\ln I_0(\eta_1\phi)$  and, hence, are neglected.
 
 The action (\ref{s-phi-small}) describes $2^{\rm nd}$ order phase transition  if the parameters satisfy
 \begin{equation}
 	\eta_0-\eta_1=1 \ .  \label{cond-1}
 \end{equation}
In dimensional units, the critical condition (\ref{cond-1})  determines a critical line as a circle of the radius $g_c$: $g^2+J^2 = g_c^2 
	 $. 
 However, the initial condition on small $\eta_1\ll \sqrt\gamma$ leaves  narrow regions from this circle 
 on a phase diagram: TCM behavior is realized when $g=g_c$ and  anti-resonant coupling    is limited  as $J  \ll \Delta$, where 
 \begin{equation}
 	  \Delta= \sqrt{\frac{\omega T}{N}} \ . \label{Delta}
 \end{equation}
 For anti-TCM  a dual condition   holds:  $g  \ll \Delta $ and $J = g_c$.

 The energy scale $\Delta$  plays  a central role in our solution, because it determines an area inside the critical  region where non-resonant terms are irrelevant, and  also the width of the Ginzburg-Levanyuk fluctuational region of the normal-to-superradiant transition.   
 The fact that this width is also equal to $\Delta$ is found from a  matching condition $\gamma^{-1/2}\sim\phi_{\rm min} $.   Here  we match     the   width of the Gaussian integrand, $\gamma^{-1/2}$, and    the value of $\phi=\phi_{\rm min}$ where the functional~(\ref{s-phi-small}) has minimum and corresponds to the superradiant phase with $\eta_0-\eta_1>1$,
  \begin{equation}
  \phi_{\rm min} = \frac{\eta_0-\eta_1-1}{2\gamma}  \ .
\end{equation} 
We note that  a non-zero $\Delta$  appears only when $T$ and $1/N$  are simultaneously non-zero.
 It  vanishes as  $\Delta\propto N^{-1/2} $ in the thermodynamic limit.

Let us find fluctuational parameters for these two areas (TCM and anti-TCM) of the critical region.
As  long as   $\eta_1\ll \sqrt\gamma$, one has   $\eta_0\approx 1$ according to (\ref{cond-1}) and 
 the action (\ref{s-phi-small}) is reduced to   
 Gaussian  form  \begin{equation}S_\phi= \gamma \phi^2 \ . \label{s-quadratic}
 	\end{equation}
 From (\ref{s-quadratic})  we find  the average photon number (it is illustrated in Fig.~\ref{potential}~(a) for Goldstone potential);   in dimensional units it reads
 \begin{equation} \langle \Phi \rangle_{\rm TC}  =\frac{1}{\sqrt\pi}  \mathcal{Q}  
 \ ,
 	\label{n-ph-I} 
 \end{equation}
 where we introduce  scaling function 
  \begin{equation}    \mathcal{Q}= \frac{1}{\beta \omega \sqrt{\gamma}}=\left[  \frac{NT\epsilon}{   \omega^2} \right]^{1/2} 
 	\label{Q} 
 \end{equation}
  that determines the  photon number, fluctuations and Fano factors. The assumption that the leading part in photon number is provided  by the condensate means $\mathcal Q \gg 1$. This gives a modification of low temperature constraint in the following form
  \begin{equation}
  \epsilon\gg T \gg T^*_{\rm TC} \ ,
  \end{equation}  where the minimal temperature is
 \begin{equation} 
 	T^*_{\rm TC}= \frac{\omega^2}{N \epsilon} \ . 
	\end{equation}
	In other words, our theory based on condensate functional is applicable if a temperature is above $T^*$. For $T$ less than $T^*$ one has  to   include higher orders in the logarithm expansion of $S_{\rm eff}$, see Eq. (\ref{s_eff_1}).
The fact that we can not go down to arbitrary low temperatures is  discussed  in  Section \ref{Minimal_temperature} in more details.

\begin{figure}[htp]
	\center{\includegraphics[width=0.99\linewidth]{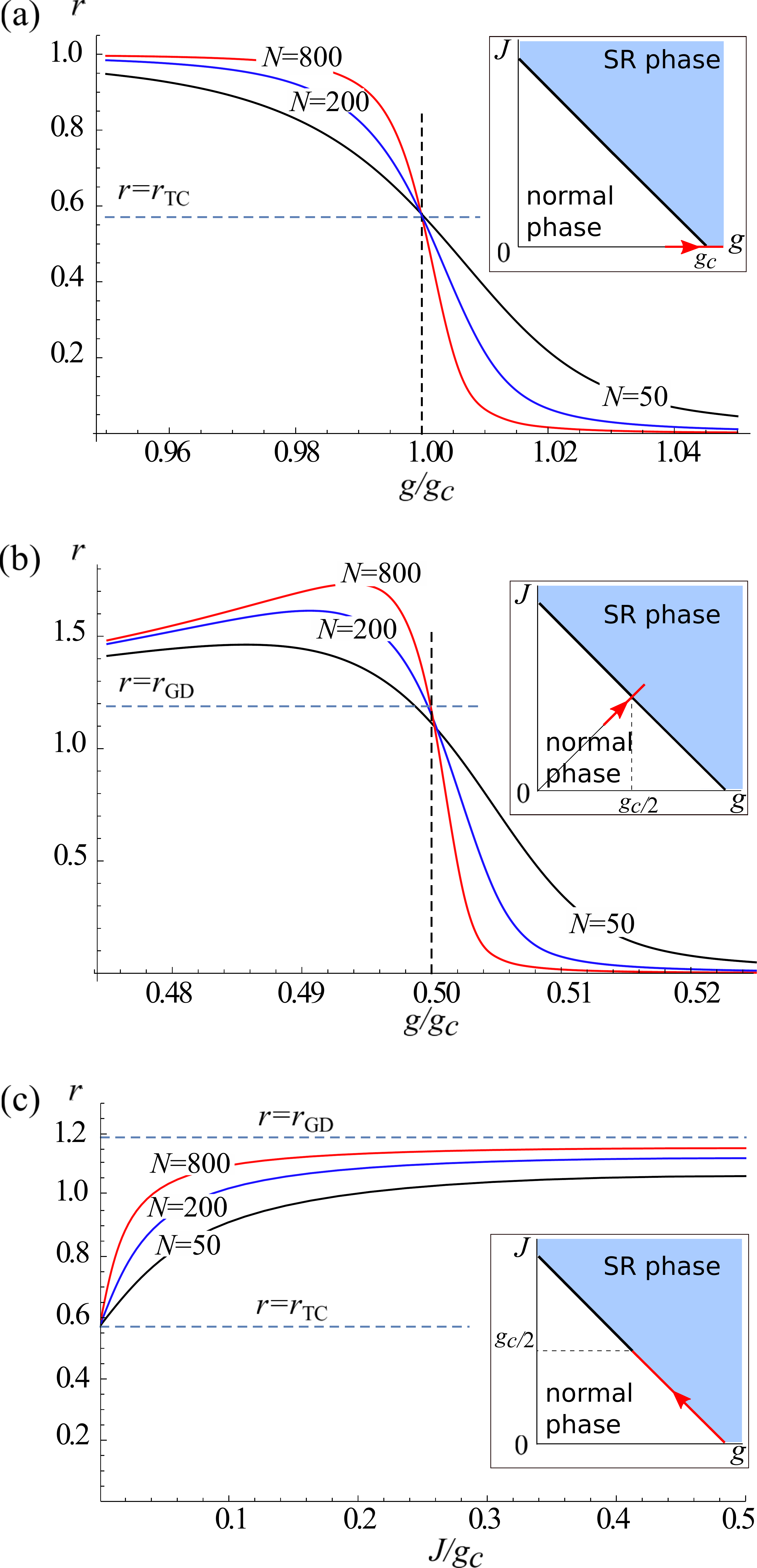}
	}
	\caption{ 
	Relative fluctuations $r$ of photon condensate near superradiant phase transition as a function of   interaction strengths. Insets: red cut in schematic phase diagram shows  a curve in $g$ and $J$ parameter space   for which $r$ dependence is plotted. Parameters   used in calculations: $\beta\omega=\beta\epsilon=100$, three curves in each plot  correspond to  $N=$ 50 (black),  200 (blue),   800 (red). (a): Regime of TCM; crossing of   curves $r(g)$ for different $N$ near the critical point $g/g_c=1$, where relative fluctuations take universal value   $r_{\rm TC}=\frac{\pi}{2}-1\approx 0.57080$.    (b): Regime of GDM, the symmetric case of $J=g$; crossing of $r(g)$   at the critical interaction $g/g_c=\frac{1}{2}$, where  relative fluctuations take   universal value   $r_{\rm GD}=4 \frac{ \Gamma^2(5/4)  }{ \Gamma^2(3/4)}-1\approx 1.18844$.   (c): Dependence of $r$ on anti-resonant coupling strength  $0{<}J{<}g_c/2$ along the critical line $J+g=g_c$. Here $J=0$ and $J/g_c=\frac{1}{2}$ correspond to TCM and symmetric Dicke model, respectively, and $r$ evolves between two asymptotical values of  $ r_{\rm TC}$ and $ r_{\rm GD}$. } \label{r-fig}
\end{figure}

 Calculations with the  Gaussian action (\ref{s-quadratic})       give the following result for relative  fluctuations
 \begin{equation} r_{\rm TC} = \frac{\pi}{2}-1\approx 0.57080  \ , \label{rc-Dicke} 
  \end{equation}            
 and the universal ratio for Fano factor
 \begin{equation}
  F_{\rm TC} /\mathcal{Q}=\left(\frac{\sqrt\pi}{2}-\frac{1}{\sqrt\pi}\right)  \approx 0.32204  \label{F-Dicke}    
 \end{equation}
   (they were previously  obtained in Ref.~\cite{PhysRevA.99.063821}). 
  In Fig.~\ref{r-fig} (b) $r(g)$ dependence  near the critical point $g=g_c$ is plotted for   $N=50$ (black curve), $N=200$ (blue curve) and $N=800$ (red curve). The crossing of all the curves in the same point $r(g_c)=r_{\rm TC}$ demonstrates the universality of this parameter.
   
 The squeezing parameters show that the condensate is slightly squeezed in $p$-direction for a non-zero  $J$:
 \begin{eqnarray} \delta x_{\rm TC}  = \sqrt{ \frac{ \langle \Phi \rangle_{\rm TC}}{2} }
 	\left(1+ \frac{J}{4}\sqrt{\frac{\pi N }{T\omega}} \right) \ , \nonumber  \\
	 \delta p_{\rm TC}  =  \sqrt{\frac{ \langle \Phi \rangle_{\rm TC}}{2}}
 	\left(1- \frac{J}{4}\sqrt{\frac{\pi N }{T\omega}} \right) \ .  \label{sq-I} 
 \end{eqnarray}
 One obtains a small correction to 1/2 in fluctuations of the phase:
 $ \langle \sin^2\varphi\rangle_{\rm TC} =1/2- \frac{\eta_1}{4\sqrt{\pi \gamma}} .  
 $
 In the dimensional  units, this is
 \begin{equation} \langle \sin^2\varphi\rangle _{\rm TC}=\frac{1}{2}- \frac{	J }{2\sqrt{\pi T \omega}}   \ .  \label{p_y^2-Dicke} 
 \end{equation}
 For a definiteness, we  assumed  TCM limit in the derivation of  (\ref{p_y^2-Dicke}, \ref{sq-I}); for anti-TCM results are the same as above with   $J$ replaced by $ g$.
 A comparison of (\ref{sq-I}) and (\ref{p_y^2-Dicke}) shows that the difference between $\delta p_{\rm TC}$ and the factorized product $(\langle \Phi \rangle_{\rm TC} \langle \sin^2\varphi\rangle_{\rm TC} )^{\frac{1}{2}}$  appears   in   linear by $J$ term.

\subsubsection{Crossover regime} 
\label{Crossover} 
Here we analyze a special case of  $
\eta_1 = 2\sqrt{\gamma} $ when the system is near the superradiant phase transition.  According to (\ref{cond-1}), it means that $\eta_0=1-2\sqrt{\gamma} $.  This point corresponds to a case  when anti-resonant interaction $J$ approaches $ \Delta$ and  TCM-to-GDM crossover  occurs.

The quadratic part in the functional (\ref{s-phi-small}) vanishes at the crossover  and  it becomes  quartic, 
\begin{equation}S_\phi= \frac{\gamma}{4} \phi^4 \  . \label{s-quartic}
\end{equation}
Remarkably,    scaling functions  of     photon number and their fluctuations  found from the quartic action  are    the same  as for TCM: $\langle \Phi \rangle^2\sim \sqrt{\langle\!\langle \Phi^2  \rangle\!\rangle}\sim\mathcal{Q} $. This is an important result we learn from the action (\ref{s-quartic}). Another one is that universal behavior of fluctuations, in other words, the parameter $r$  changes at TCM-to-GDM crossover; it   follows    from the change  of $S_\phi$ to   quartic structure.  This is due to the different prefactors in front of $\mathcal{Q}$ in   expressions for $\langle \Phi \rangle  $ and $ \sqrt{\langle\!\langle \Phi^2  \rangle\!\rangle}$ calculated with (\ref{s-quartic}). Note, that accurate calculations of $r$ and $F$  require an inclusion of higher order terms in  (\ref{s-quartic}), because   $z$ is  close to unity in the initial expansion  (\ref{LogI0-small}) for $\ln I_0(z)$ at the crossover point. 

 This asymptotic behavior of $S_\phi$ is valid for small deviation  of $\eta_1$ from  $\eta_1=2\sqrt{\gamma}$    restricted  by the condition   $ |2\sqrt{\gamma}-\eta_1|\ll \sqrt{\gamma}    
$.   In dimensional units such a condition is equivalent to
\begin{equation} 
\left|\Delta -  J \right|\ll \Delta  \	. 
	  \label{cond-quartic-3} 
\end{equation}
  Of course, in thermodynamic limit we have $\Delta=0$ and    a smooth transition between $J=0$ and $J\neq 0$ models no longer exists. A condition similar to (\ref{cond-quartic-3})  holds for $g$ in the opposite limit of anti-TCM.

In our case, when $N$ and $T$   are  finite, the matching  condition   $\Delta\sim J$ at the TCM-to-GDM crossover  can be inverted. One finds a character crossover  temperature   
 \begin{equation}
 	T_{\rm crs}=  {\frac{J^2 N}{\omega}} \ . \label{Tcrs}
 \end{equation}
 The presence of  weak anti-resonant  interaction $J$ in $\hat H$ becomes irrelevant for $T>T_{\rm crs}$ and the system behaves according to TCM model with the effective Goldstone   functional (the same  logic is applied for  anti-TCM).

\subsubsection{Generalized Dicke model regime. Squeezing}\label{Full_Dicke}

The third    type  of universal behavior at the critical region is provided by $z\gg 1$ and corresponds to $\mathbb{Z}_2$-GDM. 
Let us  come back to the action (\ref{s-phi}) and approximate the modified Bessel function   by  its large argument asymptotics, \begin{equation}I_0(z\gg 1)=\frac{1}{\sqrt{2\pi z}}e^{ z} \ .
	\end{equation} 
As a result, we obtain for the effective action
\begin{equation}
	S_\phi = (1-\eta_0 ) \phi+ \gamma \eta_0 \phi^2  + \frac{1}{2}\ln\phi \ .  \label{s-phi-large}
\end{equation}
  We note that the logarithmic divergence in (\ref{s-phi-large}) at $\phi=0$ provides  $ \frac{1}{\sqrt\phi}$ singularity in a path integral if  the logarithm is descended from the exponent. However, any of the moments   $\langle \phi^k\rangle$ with $k\geq 0$  are integrable:   
  \begin{equation}
  	\langle \phi^k\rangle_{\rm GD}\sim\int\limits_0^\infty\phi^k \frac{1}{\sqrt\phi}  e^{-(1-\eta_0 ) \phi- \gamma \eta_0 \phi^2} d\phi \ . \label{phi-k}
  \end{equation}
The representation (\ref{phi-k}) of the path integral  restores a quadratic structure of the action, similarly to TCM, with the difference that   the   multiplier $ \frac{1}{\sqrt\phi} $   appears  in front of the exponent. This results in quantitative distinctions of fluctuational behavior from that   studied  for   TCM regime.

\begin{figure}[htp]
	\center{\includegraphics[width=0.92\linewidth]{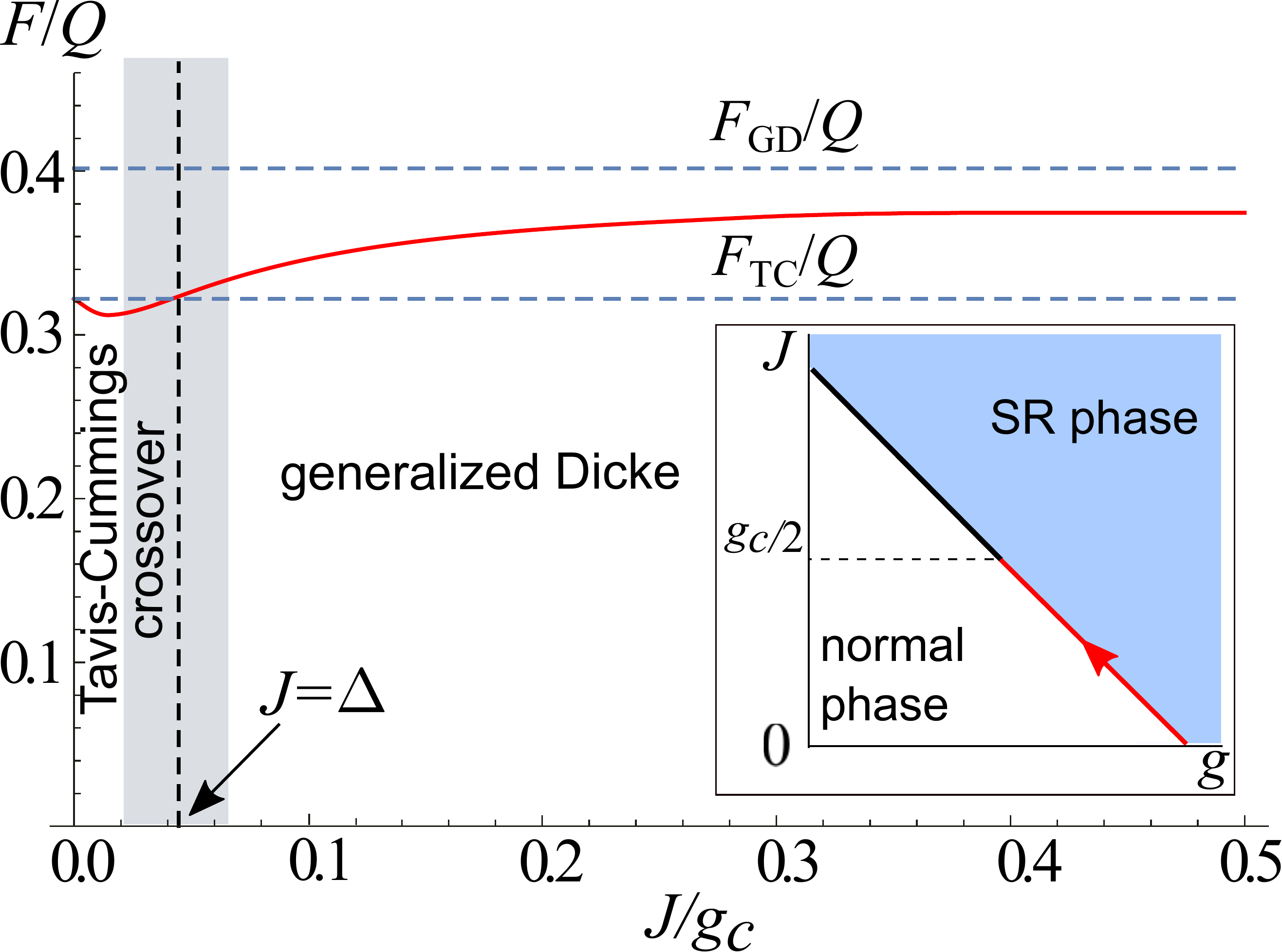}	}
	\caption{ Fano factor ratio $F/\mathcal{Q} $ as a function of anti-resonant interaction $J$. The dependence is plotted for the half of the critical line, shown as red cut in the inset, where $F$ is parametrized by anti-resonant coupling strength  $0{<}J{<}g_c/2$.   $F/\mathcal{Q} $ evolves between two universal  values of  $F_{\rm TC}/\mathcal{Q}{=}\frac{\sqrt\pi}{2}{-}\frac{1}{\sqrt\pi}\approx 0.32204$ and $ F_{\rm GD}/\mathcal{Q}{=}\frac{   \Gamma  \left(5/4\right)   }{ \Gamma  \left(3/4\right)} {-} \frac{  \Gamma  \left(3/4\right)  }{4\Gamma  \left(5/4\right) } \approx 0.40168  $    (shown as dashed blue lines). The vertical dashed line at $J=\Delta$   and shaded area nearby   separate  TCM  and GDM regimes. }   \label{fig-F}
\end{figure}

Let us determine the superradiant phase transition  point in GDM via the exponent in  representation (\ref{phi-k}). It  follows from cancelling of the linear in $\phi$ term: 
\begin{equation}
	\eta_0 =1 \ . \label{crit-line-0}
	\end{equation} 
	This critical point is  different from  that derived for TCM (\ref{cond-1}).  At the critical point, another  dimensionless interaction parameter belongs to  the domain 
\begin{equation}\sqrt{\gamma}\ll \eta_1 
< \frac{1}{2} \ , 
\end{equation}
where the upper bound $\eta_1=1/2$ corresponds to the  critical  point of the symmetric Dicke model with  $g=J=g_c/2$.
Reformulating  (\ref{crit-line-0}) through $g$ and $J$, we arrive at the critical line previously derived in Refs.~\cite{ALCALDE20113385,Alcalde_2007}: \begin{equation}g +J  =g_c \ .
	     \label{qpt-Rabi}
\end{equation}
According to the initial requirement of $\eta_1 \gg  
\sqrt\gamma$, the condition (\ref{qpt-Rabi}) is complemented by  constraints $g\gg \Delta$ and $J\gg \Delta$ which means that  we are beyond TCM-to-GDM crossover.  If one moves along $J=g$ direction in phase diagram, then the width of the fluctuational region of the normal-to-superradiant transition is equal to  $\Delta$,  the same as in the case of TCM.

The photon number in GDM regime is
 \begin{equation}
	\langle\Phi\rangle_{\rm GD}=  \frac{\Gamma(3/4)}{4\Gamma(5/4)}  \mathcal{Q} \ 
	\end{equation}
with the same scaling function $\mathcal{Q}$  as that in $U(1)$ case  (\ref{n-ph-I}) but different  from $\frac{1}{\sqrt\pi}\approx 0.56419$ prefactor of $\frac{\Gamma(3/4)}{4\Gamma(5/4)} \approx 0.33799 $. Here $\Gamma(y)$ is Euler gamma function which appears due to $\frac{1}{\sqrt\phi} $ term in path integrals (\ref{phi-k}). The relative  fluctuations parameter is not $r=\frac{\pi}{2}-1\approx 0.57080$ anymore (see Eq.~(\ref{rc-Dicke})),  it takes another universal value  
 \begin{equation}
r_{\rm GD} = 4 \frac{ \Gamma^2(5/4)  }{ \Gamma^2(3/4)}-1 \approx 1.18844  \ .\label{r-III}
\end{equation}
  This result means that relative fluctuations of the condensate are increased by the anti-resonant interaction channel.
The Fano factor, as in previous case (\ref{F-Dicke}), also scales with $\mathcal Q$, however, the prefactor in front is different  and we find another universal ratio:
 \begin{equation}
	F_{\rm GD}/\mathcal{Q} =  \frac{ 4 \Gamma^2 \left(5/4\right) - \Gamma^2 \left(3/4\right)  }{4\Gamma  \left(5/4\right)\Gamma  \left(3/4\right)}  
	\approx 0.40168	\ . \label{F-GD}
\end{equation}
 This ratio is greater than $F_{\rm TC} /\mathcal{Q}\approx 0.32204$ found for  TCM which indicates that the photon bunching in the condensate becomes stronger. 
 
 In Fig.~\ref{r-fig} (b) a dependence  of $r$  
 for symmetric Dicke model near critical point $g=J=g_c/2$ is plotted (see   red cut in the inset). Three curves  for $N=50$ (black), $200$ (blue) and $800$ (red) show that if $N$ is  increased, then   $r$ becomes closer to the   universal value  $r_{\rm GD}$ at the critical point $g_c/2$.
In Fig.~\ref{r-fig}(c) we demonstrate a behavior of  $r$ along the critical line (see red cut in the  inset). It grows from one universal value, $r_{\rm TC}$, to another, $r_{\rm GD}$, when  $J$ is increased from $0$ to $g_c/2$. Slopes of the curves near $J=0$ increase with $N$.  This means that  TCM and  crossover sectors, determined by $J\sim\Delta$, become  more  narrow,   because of   $ \Delta \propto N^{-1/2}$, and the curves  approach    to  the asymptotical value of $r_{\rm GD}$.
A dependence of the Fano factor ratio $F/\mathcal{Q}$ along the critical line is shown in Fig.~\ref{fig-F}  for $N=50$ and $\beta\omega=\beta\epsilon=10$. It changes from one universal ratio, $F_{\rm TC}/\mathcal{Q}$ to another, $ F_{\rm GD}/\mathcal{Q}$, shown as dashed blue lines.  Interestingly, that dependence on $J$ is not monotonous in TCM sector. Gray area located at $J=\Delta$ stands for the crossover region.

In a calculation of squeezing through (\ref{sq-x-p}) we approximate  the integrand by leading order terms for large $z\gg 1$ as $\big(1+ \frac{I_1(z)}{I_0(z)}\big)\approx 2$ and $\big(1- \frac{I_1(z)}{I_0(z)}\big)\approx \frac{1}{2z}$. 
As follows from physical grounds there is no   squeezing of electric component of the photon field (squeezing along $x$). The leading contribution $\delta x\approx \sqrt{\langle\Phi\rangle _{\rm GD}}$,  independent on couplings for any position in the  critical line $g+J=g_c$. A dependence on $J$ appears in higher-order correction: 
\begin{equation}\delta x_{\rm GD}=
	\sqrt{\langle\Phi\rangle_{\rm GD}}
	   \left(1-\frac{\Gamma(5/4)\omega\sqrt{\epsilon T}}{4\Gamma(3/4) \sqrt{N} g J}  \right) \ . \label{sq-x-III} 
\end{equation} 
The constraint $J\gg \Delta$ prevents a divergency of the correction for small $J$. At the TCM-to-GDM crossover,    where $J\sim \Delta$, the result (\ref{sq-x-III}) matches with that obtained for TCM (\ref{sq-I}).
For magnetic field squeezing ($p$-direction) we find that the result is totally different from that in TCM case:
\begin{equation}  \delta p_{\rm GD}  =   \frac{1}{2\sqrt{\beta \omega \eta_1}}=\left[\frac{T\epsilon}{8  g J}\right]^{1/2}  \ ,  \label{sq-III} 
\end{equation}
 namely, the temperature scaling  changes from   $ \delta p_{\rm TC}\propto T^{1/4}$ to  $
\delta p_{\rm GD}\propto T^{1/2}$.
It is important, that  $\delta p $ is independent on $\gamma$  in our limit of $\eta_1
\gg \sqrt\gamma$. It is reflected in the absence of $N$ in (\ref{sq-III}), in contrast to $\delta x_{\rm GD}\sim N^{1/4}$.
It follows from (\ref{sq-III}) that the photon condensate is squeezed in $p$-direction, i.e. $\delta p_{\rm GD} <1/2$, when 
  $J> J_{\rm sq}  $ and the threshold coupling is  \begin{equation}
	  J_{\rm sq}=\frac{\sqrt{\epsilon/\omega}}{2}T \ . \label{J_sq}
\end{equation}
Thus,  GDM sector of the critical region  has both    non-squeezed  and  squeezed phases of the photon condensate. 
An inversion of (\ref{J_sq}) provides a temperature scale   
\begin{equation}T_{\rm sq}=2J \sqrt{\frac{\omega}{\epsilon}} \end{equation} 
below which, $T<T_{\rm sq}$, a non-zero anti-resonant interaction $J$ results in the squeezing.
 Note that according to (\ref{sq-III}), the maximally possible squeezing,
 \begin{equation}
 	\delta p_{\rm max} =\sqrt{\frac{T}{2\omega} } \ , \label{p_max}
 \end{equation}
 	   appears at the  symmetric point of $g=J=g_c/2
 $  (it is illustrated in  Fig.~\ref{potential}~(b) for non-Goldstone potential).

Results for $\delta p$   are presented  in Fig.~\ref{fig-sq} where vertical axis is anti-resonant interaction $0<J<g_c$ which determines a position on the critical region (red cut in the inset), and the horizontal axis is the temperature. Here, the red cut  covers the entire critical region, from TCM to anti-TCM sectors.  Dashed curves are determined by relations $J=\Delta(T)$ and $J=g_c-\Delta(T)$, they indicate for positions of crossovers into GDM sector. Blue lines are determined by  relations $J=J_{\rm sq}(T)$ and $J=g_c-J_{\rm sq}(T)$, they are boundaries between    squeezed and non-squeezed condensates.

The last remark  in this Section concerns a correlation  between    dynamics of condensate's magnitude and phase. We analyze it via the ratio $\alpha$ of squeezing $\delta p$ and a corresponding mean-field-like factorized form:
 \begin{equation}
 	\alpha=\frac{ \sqrt{\langle \Phi     \sin^2\varphi\rangle } }{\sqrt{\langle \Phi \rangle_{\rm GD}} \sqrt{ \langle \sin^2\varphi\rangle_{\rm GD} }} \ , 
 \end{equation}
  If    fluctuations of $\Phi$ and $\varphi$  are decoupled from each other, then $\alpha=1$. Correlations between of them   lead to a decrease of $\alpha$. The squeezing  and $\langle \Phi \rangle$ which enter the expression for $\alpha$ were found above;  in a calculation of phase fluctuations ${ \langle \sin^2\varphi\rangle }$ we use a lower cut-off at $\phi\sim \eta_1^{-1}$ in the numerator of (\ref{sq-x-p}) and obtain:
\begin{equation} \langle \sin^2\varphi\rangle_{\rm GD} = \frac{(\gamma/\eta_1^2)^{1/4}}{4\Gamma(5/4)}  \ .  \label{sin2-III} 
\end{equation}
 Hence, the result for squeezed phase where $g\sim J$ is 
 \begin{equation}
\alpha \sim \left[\frac{\gamma}{\eta_1^2} \right]^{1/8}\sim \left[\frac{T}{N\epsilon} \right]^{1/8}
\ . \label{alpha}
\end{equation}
The ratio   is vanishing in a large $N$ limit as $\alpha\propto N^{-1/8} $, however,  the  decay  is rather slow. 
  This means that a correlation between fluctuations of the phase and magnitude in the  squeezed    condensate  becomes significant at large $N$. The vanishing $\alpha$ means that these fluctuations can not be decoupled by a mean-field.

\subsection{Minimal temperature}
\label{Minimal_temperature}
We are back to the issue on the applicability of  our effective theory for $\psi_0$ and derive here  the respective minimal temperature scale  $T^*$. 
At this point, we give the exact definition for the photon number which involves the occupation number $\delta n$ of all non-zero modes  
\begin{equation}	\langle \hat a^\dagger \hat a\rangle_\beta =\langle\Phi\rangle+\delta n-\frac{1}{2} \ , \quad \delta n=\sum\limits_{n\neq 0} \langle \bar \psi_n  \psi_n\rangle. \label{Nph-2}
	\end{equation}
	 The term $-1/2$ is due to commutation relations and is not important here.
The central  assumption of this work is that the leading contributions in the  thermodynamic averages  (\ref{aa}) are given    by  $\langle \Phi \rangle$.   This means  that   \begin{equation}
\delta n\ll \langle \Phi \rangle \   \label{cond-delta_n}
\end{equation}
providing  a criterion on the smallest scale $T^*$.  As   shown above, the critical scaling of photon number 
remains invariant  as  $\langle \Phi \rangle 
\sim 
\mathcal{Q}$  in GDM, (anti-) and TCM,  and the crossover regimes.   Let us analyze $\delta n$ for these cases. To do that we perform Gaussian integration with the action for non-zero $S_{\rm fl}[\bar\psi_{n\neq 0}; \psi_{n\neq 0}]$.  The  result for $\delta n$  at arbitrary $\epsilon$ and $\omega$ is cumbersome, however, for  $\epsilon=\omega$  the following  compact   form is obtained for critical line $J=\omega-g$:
\begin{equation}   \delta n =   \frac{(g+\omega)\coth \frac{\sqrt{ g \omega   }}{  T}}{8\sqrt{ g \omega   } }+\frac {\omega-g} {24T} - \frac{T}{8}\left(\frac{1}{g}+\frac{1}{\omega} \right) \ . \label{delta-n}
	\end{equation}
	We focus on the case $\omega=\epsilon$ which is rather representative and allows one to find character values of $T^*$ and its scaling with $N$.
 In contrast to leading order term $\langle\Phi\rangle$ proportional to $\mathcal Q$, temperature scaling of $\delta n$  is   sensitive to the position in the critical region. 
 There are three limits of  interest (hereafter $g_c=\omega$ and, as usual, $T\ll \omega$).  The first one is given by RWA, where $J\ll\Delta$ and $g\approx g_c$. The correction and minimal temperature, that follow  from   (\ref{cond-delta_n}) and (\ref{delta-n}), read as 
  \begin{equation} \delta n_{\rm TC} = \frac{1}{4}+O(T/\omega) \ , \quad T^*_{\rm TC}=\frac{\omega}{N}  \ . \label{delta-n-rwa}
	\end{equation}
 At this point, we reproduce the result of Ref.~\cite{PhysRevA.99.063821}  on the applicability of the $S[\Phi,\varphi]$
 \begin{equation}  \omega\gg T \gg T^*_{\rm TC}  \ . \label{cond-temp}
 \end{equation}
   In GDM regime, the leading part in $\delta n$ grows with $J$ approaching to $\omega$ as
    \begin{equation} \delta n_{\rm GD}(J) =\frac{1}{8}\sqrt{\frac{1}{1-J/\omega}} \ , 
     \label{delta-n-dicke}
 \end{equation}
where $ \omega-J\gg \Delta$.
 The minimal temperature also grows inverse proportional to $J$  
    \begin{equation}
   	  T^*_{\rm GD} (J)  =\frac{\omega^2}{N(\omega-J)}  \ . \label{t-min-dicke}
\end{equation}

A strikingly different result for the minimal temperature is found for anti-TCM domain, where  $\omega-J\sim \Delta$. In this case the leading part in $\delta n$ is given by    \begin{equation}
\delta n_{\rm antiTC}  =\frac{ \omega}{24 T} \ . 
  \end{equation} 
  This   provides  the distinct scaling  law for $N$:
 \begin{equation}      T^*_{\rm antiTC}  =\frac{\omega}{N^{1/3}}  \ . \label{t-min-crt}
\end{equation}
Thus, we obtain   $T^*_{\rm antiTC} \gg T^*_{\rm TC}$. It means that   fluctuations above the condensate in anti-TCM are stronger than that in TCM. The difference in the scaling laws   is a manifestation of that fact that these models are not dual to each other.  

\begin{figure}[htp]
	\center{\includegraphics[width=0.99\linewidth]{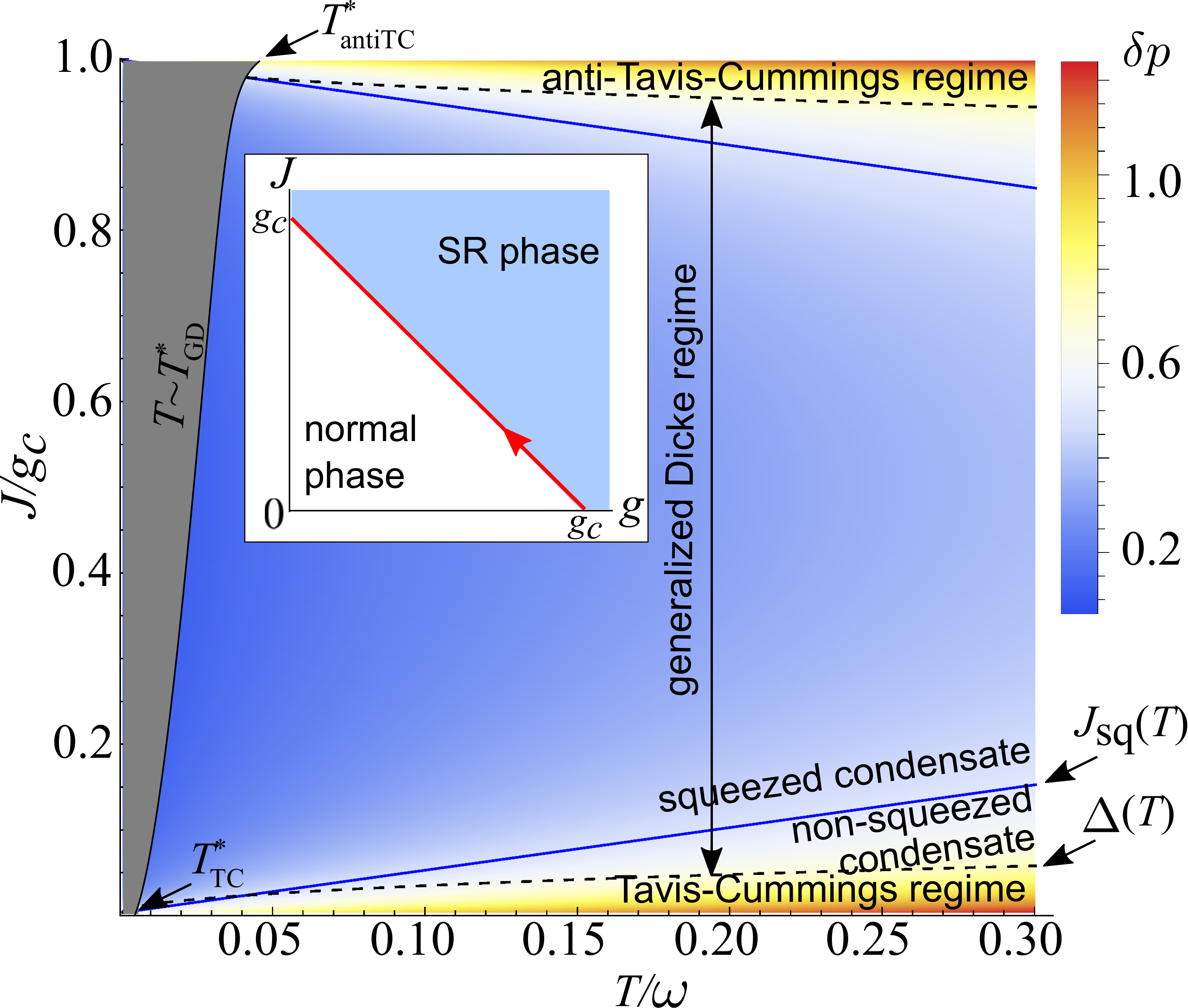}	}
	\caption{Different regimes for   $\delta p$ squeezing in the critical region as a function of  temperature $T$. Parameters $g$ and  $J$ correspond to the red cut in the inset which covers the critical region entirely; the plot is parametrized by $J$ (vertical axis).  Blue lines $J=J_{\rm sq}(T)$  and $J=g_c-J_{\rm sq}(T)$ are  boundaries    between squeezed and non-squeezed condensates. Dashed curves corresponds to crossovers between (anti-) and TCM   and GDM regimes.  The gray sector is the part of phase diagram where quantum fluctuations due to non-zero Matsubara modes are important. Here, our solution based on the effective functional for condensate modes only is not strict. The edge curve of the gray sector demonstrates schematically a dependence of minimal temperature on $J$. Its value is minimal near    $J=0$, where $T=T^*_{\rm TC}$, then increases with $J$ to $T\sim T^*_{\rm GD}$ and approach to $T=T^*_{\rm antiTC}$  in the anti-TCM sector.  } \label{fig-sq}
\end{figure}

Numerical solution $\delta n  =\mathcal Q$ gives a typical dependence of $T^*$ on $J$. This solution is presented   in Fig.~\ref{fig-sq}  as the edge of the gray sector, where $T^*$ increases with $J$   according to the above   analysis.

\subsection{Zero-temperature limit for (anti-) and Tavis-Cummings models} \label{Bethe_ansatz}

It is of   interest to analyze in  more detail zero-temperature properties of the system under the consideration. Both TCM and anti-TCM limits are exactly solvable using Bethe ansatz even if inhomogeneous broadening is present in the system. Let us first consider TCM regime. As it was already mentioned above, in this limit the excitation number is a good quantum number, since its operator commutes with the Hamiltonian. Hence, there exist sectors with different excitation numbers $N_{\rm ex}$ or with different excitation densities $\rho=N_{\rm ex}/N$, provided the thermodynamical limit $N \rightarrow \infty$ is considered. The leading in $1/N$ contribution to the ground state energy density at given $\rho$ is \cite{Pogosov_2017}
\begin{multline}
E_{\rm gr}(\rho) / N = \frac{1}{2}\left(\varepsilon-\sqrt{(\varepsilon-\lambda)^2+\xi^2}\right)+ \\ + \lambda \left(\rho-\frac{1}{2}\right)+\frac{\xi^2}{4}\left(\omega-\lambda\right),
\label{Bethe}
\end{multline}
where both parameters $\xi$ and $\lambda$ are determined by conditions $\partial E_{\rm gr}(\rho) / \partial \xi=0$ and $\partial E_{\rm gr}(\rho) / \partial \lambda = 0$. The ground state energy $E_{\rm gr}(\rho)$ is an extensive quantity. We stress that all other contributions to the energy are negligible in the limit $N \rightarrow \infty$, i.e., non-extensive, but they can be evaluated using the approach of Ref. \cite{Pogosov_2017}. At fixed $\rho$, parameters $\xi$ and $\lambda$ also determine energies of excited dressed states (TCM Hamiltonian eigenstates) given by $\sqrt{(\varepsilon-\lambda)^2+\xi^2}$. In this sense, $\xi$ and $\lambda$ are similar to the gap and chemical potential, respectively, while the mean-field treatment turns out to be exact in the thermodynamical limit due to a specific structure of the interaction term of the Hamiltonian (all-to-all interaction). Note that the mean density of photons is expressed through $\xi$ as $\xi^2/2g^2$.

The global ground state of the system is given by the minimum of $E_{\rm gr}(\rho)$ as a function of $\rho$. It is easy to find from the above two equations that, at $g$ small enough, this global minimum corresponds to the normal phase with $\rho=\xi=0$. This result agrees with the perturbation theory around the noninteracting limit $g=0$. The normal phase becomes unstable ($dE_{\rm gr}(\rho)/d\rho=0$ at $\rho=0$) at the critical coupling 
$g=g_c$ where $2^{\rm nd}$-order phase transition emerges. It is  accompanied by the appearance of both nonzero excitation and photon densities given by $\rho$  and  $\xi^2/2g^2$, respectively.

Now we discuss the anti-TCM limit. Mathematically, it can be mapped on the TCM regime by considering another vacuum state with all qubits   excited. The Hamiltonian acting on this polarized vacuum acquires an additional contribution $\varepsilon N/2$, while the excitation energies of qubits are transformed as $\varepsilon \rightarrow - \varepsilon$ 
and $\hat\sigma^+_j \rightarrow \hat\sigma_j^- $, $\hat\sigma_j^- \rightarrow \hat\sigma_j^+ $. Under such a mapping, Bethe ansatz can be applied  as well. We also should keep in mind that the normal state now corresponds to $\rho=1$ in terms of excitations of the new vacuum state. By performing the same analysis as it the case of TCM, we find that the normal state with zero photon density becomes unstable towards a superradiant phase with nonzero photon density at the same interaction constant $J=g_c$ and the transition is also of  $2^{\rm nd}$-order. This is again in agreement with the path integral treatment.

In the view of the duality between TCM and anti-TCM, the latter result may seem as rather expectable, but we would like to stress that, by its structure, the anti-resonant interaction term is quite different from the resonant one and, therefore, normal states must differ in TCM and anti-TCM limits. Indeed,  the resonant term does not change an excitation number and therefore the normal phase contains exactly zero excitations. In contrast, anti-resonant term does change an excitation number and hence photons should be present even in the normal state, as follows from the perturbation theory near $g=0$ limit. However, photon density vanishes in the thermodynamical limit, while photon number does not (this is also readily revealed using the perturbation theory). From this viewpoint, the similarity of TCM and anti-TCM is not obvious and it emerges in the thermodynamical limit only, 
 while finite-$N$ regimes must be different. The difference between anti-TCM and TCM can be also linked to the fact that anti-TCM is mapped on TCM with negative qubit excitation energies and physically the duality is not absolute since normal states correspond to different values of $\rho$. It is evident from the above considerations that finite-size corrections (in powers of $1/N$) to TCM and anti-TCM regimes 
 differ. This  latter  conclusion is justified by the results derived 
 with the use of the  path integral formalism.

 \section{Discussion } \label{Discussion}
  Let us now discuss a connection of our results to that known from some other works on superradiant QPT. 
 As was shown 
 in Refs.~\cite{PhysRevLett.90.044101, PhysRevE.67.066203},    symmetric Dicke model  
 reveals  
 signatures of  quantum chaos above the superradiant QPT  if   $N$ is finite. It was  shown 
 through  quasi-classical equations of motion 
 and also through   
 a change of the eigenvalues statistics 
 from Poissonian,  
 in the normal phase, 
 to a Wigner one, above the QPT. 
The repulsion of levels in  GDM with $g\neq J$ and an interpretation of that as   quantum chaos        was discussed   earlier, in particular, 
 in    Ref.~\cite{LEWENKOPF1991113}. In that work authors demonstrated a variety of non-regular  levels statistics for different $g/J$ ratios, however, a connection  with the superradiant transition was not discussed. 
  Leaving the issue on levels statistics    near  the critical line    $g+ J=g_c$ beyond the scope the consideration, we provided a description of the macroscopic photon condensate properties in this work implying that microscopic  dynamics can be strongly   chaotic. Our results on   universal fluctuations and field squeezing, collected in Table~\ref{table-0}, give an alternative view on ergodic dynamics of GDM. 
 \begin{table*}[ht]
\caption{Results for universal fluctuations, squeezing,    and minimal temperature.}
\centering
\begin{tabular}{ | p{0.1\linewidth}
| p{0.13\linewidth}| p{0.17\linewidth}| p{0.17\linewidth}| p{0.17\linewidth}|p{0.1\linewidth}|}
\hline
 Regime & 
 Relative fluctuations, $r$ & Fano factor ratio, $F/\mathcal{Q}$ & Coordinate squeezing, $\delta x$ & Momentum squeezing, $\delta p$  & Minimal temperature, $T^*$\\
\hline
\vspace{0.3cm}Tavis-Cummings &
\vspace{0.3cm}$$\frac{\pi}{2}-1$$ & 
 \vspace{0.2cm}$$\frac{\sqrt\pi}{2}{-}\frac{1}{\sqrt\pi} $$ &  \vspace{-0.18cm}$$\frac{1}{\sqrt{2} \pi^{1/4}}\left[ \frac{NT\epsilon}{   \omega^2} \right]^{1/4}$$  & \vspace{-0.18cm}$$\frac{1}{\sqrt{2} \pi^{1/4}}\left[ \frac{NT\epsilon}{   \omega^2} \right]^{1/4}$$ & \vspace{0.3cm}$$\frac{\omega}{N}$$  \\
\hline
 \vspace{0.3cm} Generalized Dicke & 
$$  4 \frac{ \Gamma^2(5/4)  }{ \Gamma^2(3/4)}-1$$ &  $$ \frac{ 4 \Gamma^2 \left(5/4\right) - \Gamma^2 \left(3/4\right)  }{4\Gamma  \left(5/4\right)\Gamma  \left(3/4\right)}  $$ &  \vspace{-0.1cm}$$\sqrt{\frac{\Gamma(3/4)}{4\Gamma(5/4)}  }\left[ \frac{NT\epsilon}{   \omega^2} \right]^{1/4} 
	   $$ & \vspace{0.25cm}$$ \left[\frac{T\epsilon}{8  g J}\right]^{1/2} $$& $$\frac{\omega^2}{N(\omega-J)} $$ \\
\hline
\vspace{0.3cm} anti-Tavis-Cummings& 
\vspace{0.3cm}$$\frac{\pi}{2}-1$$ & 
 \vspace{0.2cm}$$\frac{\sqrt\pi}{2}{-}\frac{1}{\sqrt\pi} $$ & 
 \vspace{-0.18cm}$$\frac{1}{\sqrt{2} \pi^{1/4}}\left[ \frac{NT\epsilon}{   \omega^2} \right]^{1/4}$$& 
 \vspace{-0.18cm}$$\frac{1}{\sqrt{2} \pi^{1/4}}\left[ \frac{NT\epsilon}{   \omega^2} \right]^{1/4}$$ & $$ \frac{\omega}{N^{1/3}}$$ \\
\hline
\end{tabular} \label{table-0}
\end{table*}
 In our approach we analyzed non-Goldstone  functional   that  depends on  two   variables, the magnitude   and   phase of the   photon condensate. 
 We showed that a coupling between fluctuations of these variables  can be reduced to an effective dissipative action~(\ref{s-phi}) for the magnitude only. (This quantity is proportional   to the  condensed photons amount.) The functional has different asymptotical behavior depending on $g$-to-$J$ ratio at the   critical region of normal-to-superradiant phase transition.

 Rather remarkable,   a structure of the functional  along $g+J=g_c$ is   such   that 
   scalings of photon number and its fluctuations remains unchanged as 
  $\langle \hat a ^\dagger \hat a  \rangle \sim \langle\!\langle \hat a ^\dagger \hat a\hat a ^\dagger \hat a\rangle\!\rangle^{1/2} \sim \mathcal{Q}   
 $ with the scaling function  $\mathcal{Q}=
   \sqrt{NT\epsilon/\omega^2}$. 
 However, a sensitivity to anti-resonant coupling appears in their  relative values and  the squeezing of magnetic filed component. The phase diagram, illustrating these sectors with different universal behaviors inside the critical region, is shown  in Fig.~\ref{fig-diagram-F}. As follows from this effective theory,   relative fluctuations of   condensed photons can take two  universal values at the critical region: $r_{\rm TC}~=~\frac{\pi}{2}-1$ and $r_{\rm GD}~=~4 \frac{ \Gamma^2(5/4)  }{ \Gamma^2(3/4)}-1$, corresponding  to TCM and GDM regimes, respectively.  Character dependences of relative fluctuations are depicted in  Figs.~\ref{r-fig} (a, b) for these two limits. The effective action  derived  allows to describe  a smooth crossover between these two regimes.    Similar  crossover is found for opposite anti-TCM limit as well. 
 The  domain of  coupling $J$, where anti-resonant terms in the Hamiltonian are irrelevant and the condensate behaves accordingly to finite $T$ TCM, is limited from above as $J\ll \Delta$ where the character energy scale  is $\Delta=\sqrt{ \omega T /N}$.
In Fig.~\ref{r-fig}(c) we showed how the relative fluctuations $r$  evolve  from $r_{\rm TC}$ to $r_{\rm GD}$   when one moves along the critical region and crosses $J= \Delta$.

 The Fano factor  found  is much greater than unity at the critical region  indicating for a photon bunching effect or, in other words, for a positive photon-photon correlation. Ratios $F/\mathcal Q$ were found to be  universal constants, again, in TCM and GDM sectors of phase diagram (their values are presented in  Table~\ref{table-0}). The crossover from one universal value, $F_{\rm TC}/\mathcal{Q}$, to another, $F_{\rm GD}/\mathcal{Q}$,  is shown in  Fig.~\ref{fig-F}. The Fano factor   increases by a number  for $J\gtrsim \Delta$; it stands for an increase of   photon-photon correlations   due to anti-resonant term in the Hamiltonian and growing   entanglement of eigenfunctions.

\begin{figure}[htp]
	\center{\includegraphics[width=0.95\linewidth]{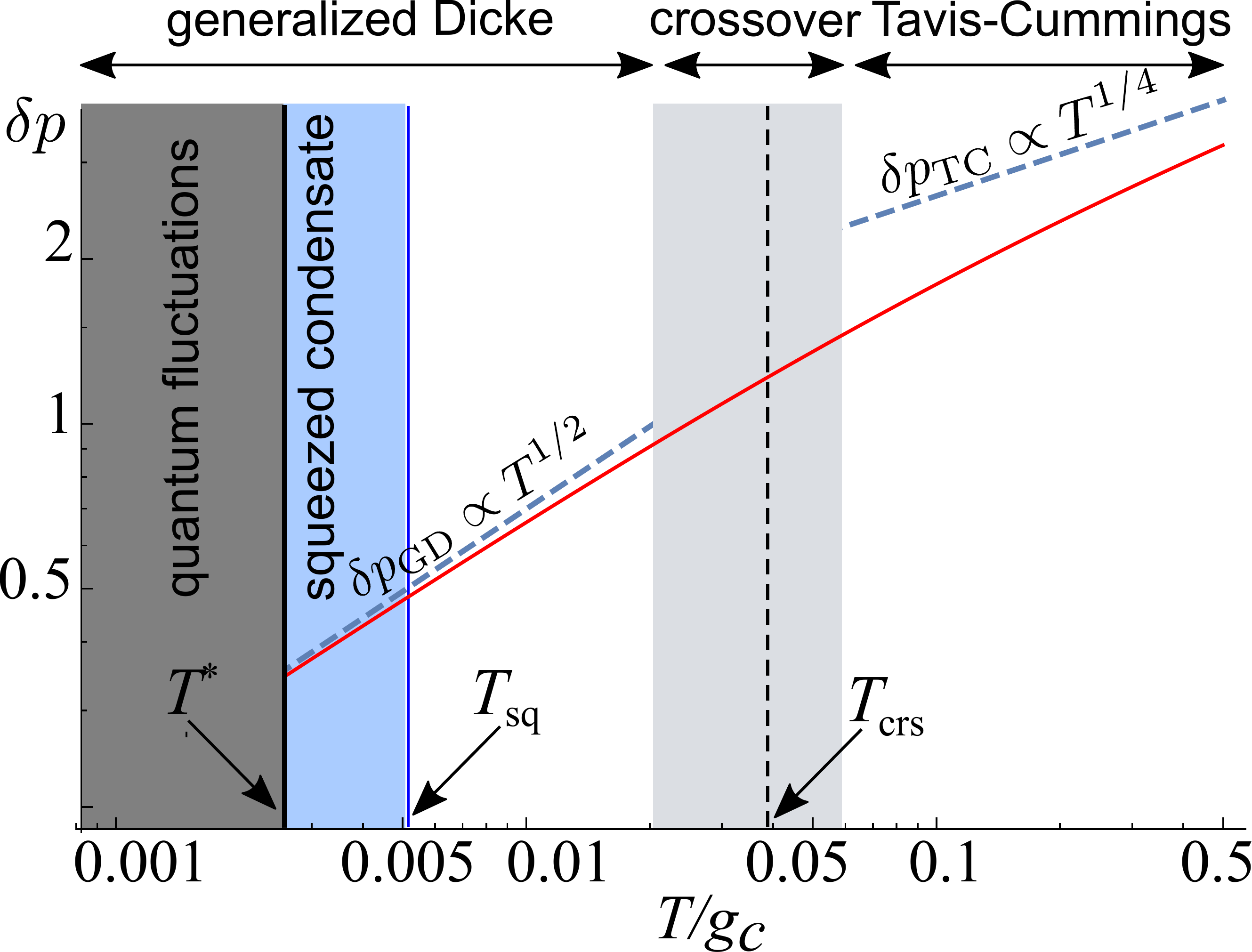}	}
	\caption{The logarithmic dependence of squeezing $\ln \delta p$ as a function of $\ln T$ (red curve) and the hierarchy of temperature  scales $T_{\rm crs}$, $T_{\rm sq}$, and $T^*$ (vertical dashed, blue and solid black lines).  
	 Universal behaviors for different $T$: TCM regime ($T>T_{\rm crs}$),  TCM-to-GDM  crossover ($T\sim T_{\rm crs}$, gray sector), GDM regime  ($T<T_{\rm crs}$).  Dashes lines: different temperature scalings of squeezing in TCM and GDM regimes. GDM sector is subdivided into three parts:   non-squeezed condensate ($T_{\rm sq}<T<T_{\rm crs}$), squeezed condensate ($T^*<T<T_{\rm sq}$, light blue sector), and quantum fluctuational dominated regime ($T<T^*$, dark gray sector). Parameters $J=0.01 g_c$, $g=0.99 g_c$, $N=100$, $\epsilon=15\omega$, and $g_c=\sqrt{\omega \epsilon}\approx 3.87298\omega$.}
		 \label{fig-hierarchy}
\end{figure}

An  important result is that momentum squeezing of the superradiant condensate  is sensitive to the anti-resonant interaction.  In $U(1)$ limit of  $J=0$,  there is no squeezing and  $\delta x=\delta p \sim \mathcal{Q}^{1/2}$. A non-zero $J$ results in a momentum squeezing, $\delta p < \delta x$. We find that it has two different temperature scalings   in TCM and in GDM sectors as $\delta p_{\rm TC}\propto T^{1/4}$ and $\delta p_{\rm GD}\propto T^{1/2}$, respectively. These asymptotics are shown as dashed lines in Fig.~\ref{fig-hierarchy} for $\delta p (T)$ plot.
 An alternative representation for $\delta p$   is   shown in $T$-$J$  phase diagram  in  Fig.~\ref{fig-sq}.  Here,    the critical region reveals  a squeezing of the photon condensate for anti-resonant coupling $J>J_{\rm sq}=T \sqrt{\epsilon/(4\omega)} $ (blue region).  Interestingly, that interaction strength $J_{\rm sq}$ scales linear with $T$ and does not depend on $N$.

 We also show positions of character temperature domains in Fig.~\ref{fig-hierarchy}.
One can see that the decrease of the temperature 
down to TCM-to-GDM crossover,  $T\sim T_{\rm crs}= N  J^2/\omega$ (light gray sector),   shows a change in the scaling of $\delta p$.  In the crossover regime  the effect of anti-resonant interaction terms becomes relevant. The universal fluctuations also change here from $r_{\rm TC}$ to $r_{\rm GD}$.
Further decrease of the temperature below $T<T_{\rm sq} = 2J\sqrt{\frac{\omega}{\epsilon}}$ shows the entrance into a  squeezed phase  of the condensate  where $\delta p < 1/2$ (light blue sector). This corresponds to  an effect of the condensate's phase fixation.  Cooling the system down to  $T^*$,  the  entrance into a quantum fluctuational dominated regime  occurs (dark gray sector). 
The   hierarchy of energy scales in the critical region  is $g_c\gg \{T, J_{\rm sq}\}\gg\{\Delta, T^*\} $.

The minimal temperature   $T^*$, that determines our effective theory as $ 
\omega\gg T\gg T^*  $, corresponds to a change of  the character  of the phase transition: it is suggested that for $T\lesssim T^*$ normal-to-superradiant fluctuational transition changes to zero-temperature QPT. There is a decrease of thermal fluctuations in the condensate mode in comparison to  quantum fluctuations encoded by a non-zero Matsubara modes. Hence,   the effective theory   allows to approach   QPT parametrically close for large $N$.

 As follows from vanishing  $\Delta\sim T^{1/2}$ at zero temperature, the critical region  shrinks to the line. Scaling behavior changes in this case as follows.  In our finite-$T$ situation, the photon number   scaling is  $ \propto N^{1/2}$.
 As shown in~\cite{refId0}  for symmetric model at $T=0$, photon number scales distinctly as $\propto N^{1/3}$ near the critical point.  This solution was obtained via Holstein-Primakoff bosonization and applies for the normal phase below QPT.  As we have already mentioned, the matching between these finite- and zero-temperature behaviors is a non-trivial issue.  Similar change in finite- and zero-temperature physics was found also for Lipkin-Meshkov-Glick finite-$N$ model  at the critical point~\cite{Wilms_2012}.

The relation between $T^*$   and $J$ is shown in phase diagram in Fig.~\ref{fig-sq} as  edge curve of the gray sector. 
It increases from   $T_{\rm TC}^* \sim \omega /N$  in TCM limit with $J=0$  to   $T_{\rm antiTC}^* \sim \omega /N^{1/3}$ in anti-TCM limit when  $J$ approaches $g_c $. The difference in the exponent ($N^{-1}$ vs $N^{-1/3}$)  is explained by different symmetries  of the respective  Hamiltonians. 
According to the above,   for arbitrary small $T$  large  $N$  exists  such that  Hilbert space dimension  of a respective   $\hat H$   compensates  exponentially small  Gibbs weights in the   density matrix $e^{- \beta \hat H }$. (As a consequence, this results in a macroscopic occupation number in the condensate with   finite-size fluctuations.) It  is supposed to be $N\gg \omega / T$ for TCM sector in Fig.~\ref{fig-sq} and more strict one, $N\gg \left(\omega/T\right)^3$, for the opposite sector of anti-TCM.   These conditions determine lower   number of two-level systems in the  ensemble  when dynamics is similar to that  in the thermodynamic limit.

\section{summary}
\label{summary}

In this work, we investigated   
an effect of anisotropic interaction between single-mode cavity and two-level systems ensemble on fluctuational properties of a photon condensate near the superradiant phase transition. Addressing to the equilibrium field theory for  generalized Dicke model, we  
focused on  a situation of simultaneously finite temperature and size of the ensemble. 
This regime was found to be more complex than the well-studied quantum phase transition at zero temperature~\cite{PhysRevLett.90.044101, PhysRevE.67.066203, PhysRevLett.108.073601,refId0,PhysRevLett.112.173601,PRL_119_220601} or thermodynamic limit at infinite ensemble's size~\cite{popov1988functional,eastham2001bose, eastham2006finite,Pogosov_2017,ALCALDE20113385,Alcalde_2007}.  
We showed that an increase of the anti-resonant coupling   changes   one critical behavior, corresponding to  Tavis-Cummings or anti-Tavis-Cummings  $U(1)$  models,  to another one, corresponding to  generalized Dicke $\mathbb Z_2$ model. 
 This transition between two fluctuational behaviors reveals a change in temperature scaling laws for squeezing parameters. The anti-resonant interaction strength, above  which the condensate becomes strongly squeezed, was determined. 
 We also found explicit expressions for other universal parameters which characterize  fluctuations; they do not depend on the temperature and ensemble's size. This is, in particular, Fano factor representing photon bunching in the condensate.  The presented study, which demonstrates a richness of the critical behavior, 
is expected to be relevant for the understanding of many-body physics of cavity quantum electrodynamics.

 The averaging with finite temperature density matrix used as a theoretical tool in our findings assumes  that the system is open. In contrast to virtual photons in a ground state at zero temperature, condensate's photons in our finite-temperature situation can be measured~\cite{Frisk-Kockum:2019aa}. This can be done by the methods employed for driven-dissipative condensate~\cite{Baumann} where the superradiant QPT
was demonstrated.  Generally, non-equilibrium conditions result in a change of universality class of the dynamics. Nonetheless,  open quantum  systems near  critical point are known to behave  effectively as equilibrium with a certain  effective temperature and obey low-frequency fluctuation-dissipation relations~\cite{PhysRevLett.110.195301,dalla2013keldysh, kirton2018introduction}.

It is suggested  that our findings might be accessible in  state-of-the-art realizations    of strongly coupled light-matter systems~\cite{Frisk-Kockum:2019aa} such as quantum metamaterials and simulators based on cold atoms, superconducting qubits, 
nitrogen-vacancy centers, and semiconductor based heterostructures. 
 A possible route can be probing of the  Fano factor through the counting of photon numbers over long times.  They are accessible through intensity (second-order)  correlation functions measurements~\cite{PhysRevLett.106.243601} or transmission of incoherent drive~\cite{PhysRevLett.107.053602} realized in photon blockade effect.  This allows one to  identify the position of the critical region of the superradiant phase transition. Then, extracting  the relative fluctuations values  with the use of the counting data one   obtains  information on the type of universal behavior and the respective symmetry of the interaction. 
 Our predictions  on   universal fluctuations and  squeezing of the photon condensate, 
 in principle, can be  measured by the dispersive readout techniques.

\section{Acknowledgments}
We thank Arkady M. Satanin, Igor S. Burmistrov, and Julien Vidal for fruitful discussions. The  work reported in Sections II~A-E and III~A.1-3 was funded by RFBR according to the research project N\textsuperscript{\underline{o}}    
19-32-80014. 
 	The work reported in Sections   III~B and III~C was financed by the Russian Science Foundation under Grant N\textsuperscript{\underline{o}}  16-12-00095.
 Yu.E.L. acknowledges  support from  RFBR project N\textsuperscript{\underline{o}}~20-02-00410. 
 W.V.P. acknowledges  support from  RFBR project N\textsuperscript{\underline{o}}~19-02-00421.
 D.S.S. acknowledges   the funding by RFBR according to  research projects N\textsuperscript{\underline{o}}~20-37-70028 and N\textsuperscript{\underline{o}}~20-52-12034.

\end{document}